\documentclass[a4paper,superscriptaddress, amsfonts, amssymb, amsmath, reprint, showkeys, nofootinbib, twocolumn, floatfix]{revtex4-1} 
\usepackage[english]{babel}
\usepackage[utf8]{inputenc}
\usepackage{amsthm}
\usepackage{mathtools}
\usepackage{physics}
\usepackage[left=23mm,right=13mm,top=35mm,columnsep=15pt]{geometry} 
\usepackage{adjustbox}
\usepackage{placeins}
\usepackage[T1]{fontenc}
\usepackage{lipsum}
\usepackage{csquotes}
\usepackage{siunitx}
\usepackage{booktabs}
\usepackage{multirow}
\usepackage{notes2bib}
\bibnotesetup{ note-name = , use-sort-key = false}
\usepackage{booktabs}
\usepackage{colortbl}
\usepackage{xcolor}
\usepackage{graphicx}
\definecolor{white}{rgb}{1.0, 1.0, 1.0}
\definecolor{beaublue}{rgb}{0.74, 0.83, 0.9}
\DeclareSIUnit\micron{\micro\metre}
\DeclareSIUnit\ms{\milli\second}

\newcommand{\be}{\begin{equation}}
\newcommand{\ee}{\end{equation}}
\newcommand{\ba}{\begin{eqnarray}}
\newcommand{\ea}{\end{eqnarray}}

\begin{document}

\title{Two-particle Interference with Double Twin-atom Beams}
    
    \author{F. Borselli}
    \author{M. Maiwöger}
    \author{T. Zhang}
    \author{P. Haslinger}
    \affiliation{Vienna Center for Quantum Science and Technology, Atominstitut, TU Wien, 
    1020 Vienna, Austria}
    
    \author{V. Mukherjee}
    \affiliation{Indian Institute of Science Education and Research, 760010 Berhampur, India}
    
    \author{A. Negretti}
    \affiliation{The Hamburg Centre for Ultrafast Imaging, Universität Hamburg, D-22761 Hamburg, Germany}
    
    \author{S. Montangero}
    \affiliation{Dipartimento di Fisica e Astronomia “G. Galilei”, Università di Padova, I-35131 Padova, Italy} \affiliation{INFN Sezione di Padova, I-35131 Padua, Italy.}
    
    \author{T. Calarco}
    \affiliation{Forschungszentrum Jülich, Wilhelm-Johnen-Straße, D-52425 Jülich, 
    and University of Cologne, Institute for Theoretical Physics, D-50937 Cologne, Germany}
    
    \author{I. Mazets}
    \affiliation{Vienna Center for Quantum Science and Technology, Atominstitut, TU Wien, 
    1020 Vienna, Austria}
    \affiliation{Research Platform MMM ``Mathematics--Magnetism--Materials'', 
       c/o Fakult{\"a}t f{\"ur} Mathematik, Universit{\"a}t Wien, 1090 Vienna, Austria}

    \author{M. Bonneau}
    \author{J. Schmiedmayer}
    \affiliation{Vienna Center for Quantum Science and Technology, Atominstitut, TU Wien, 
    1020 Vienna, Austria}

\begin{abstract}

We demonstrate a source for correlated pairs of atoms characterized by two opposite momenta and two spatial modes forming a Bell state only involving external degrees of freedom. We characterize the state of the emitted atom beams by observing strong number squeezing up to -10 dB in the correlated two-particle modes of emission. We furthermore demonstrate genuine two-particle interference in the normalized second-order correlation function $g^{(2)}$ relative to the emitted atoms.

\end{abstract}

\date{\today} 

\maketitle

Correlated and entangled pairs constitute a fundamental tool in the hands of a quantum engineer~\cite{dowling2003quantum} with a wide range of possible applications, from probing fundamental questions regarding the nature of the quantum world, to building blocks for quantum communication and quantum computers, to sensors and development of metrological devices~\cite{QuantumRoadMap}. 
Many beautiful fundamental and applied experiments have been performed with entangled pairs of photons~\cite{PanRMP2012}. 
In recent years huge progress was made in creating entangled states of massive particles, most prominently in the context of developing fundamental building blocks for quantum logic operations. The interest is also motivated by performing a Bell test using massive particles, as in spin correlations between protons~\cite{lamehi1976quantum}, electrons~\cite{hensen2015loophole}, ions~\cite{rowe2001experimental}, Josephson phase qubits~\cite{ansmann2009violation} and atoms~\cite{rosenfeld2017event}. 

The above experiments were performed for internal states and except for the proton experiment with localized systems. Here we will focus on external degrees of freedom of freely propagating pairs of atoms. The most direct way to produce them is by collisions, which can either be accomplished by collisional de-excitation in a quantum degenerate sample in an excited motional state in a trap or waveguide~\cite{bucker2011twin}, by designing the dispersion relation using a lattice~\cite{BonneauPRA2013,lopes2015atomic,dussarrat2017two} or by colliding two condensates with different momenta and looking at the scattering halo~\cite{PhysRevLett.105.190402,shin2019bell}.  

In this Letter, we present a source of double twin-atom beams (DTBs): beams of atoms emitted in pairs with opposite momenta (twin atoms) traveling in a double waveguide potential. This has two advantages. On the one hand, it forces the emission along the waveguide; hence it is more efficient than experiments where the emission happens in free space~\cite{PhysRevLett.105.190402, kofler2012einstein, lewis2015proposal}. On the other hand, the presence of two such parallel waveguides allows the possibility for the twin pair to be emitted into either the left waveguide ($L$ waveguide) or into the right waveguide ($R$ waveguide); hence an entangled state of two atoms only involving motional degrees of freedom is possible. We measure momentum correlations between the atoms in the pairs and observe a fringe pattern in the normalized second-order correlation function $g^{(2)}$ that stems from a two-particle interference phenomenon. The fundamental idea at the basis of this experiment is discussed in more detail in Ref.~\cite{bonneau2018characterizing}.

\begin{figure*}[htb]
        \centering
        \includegraphics[width=0.8\textwidth]{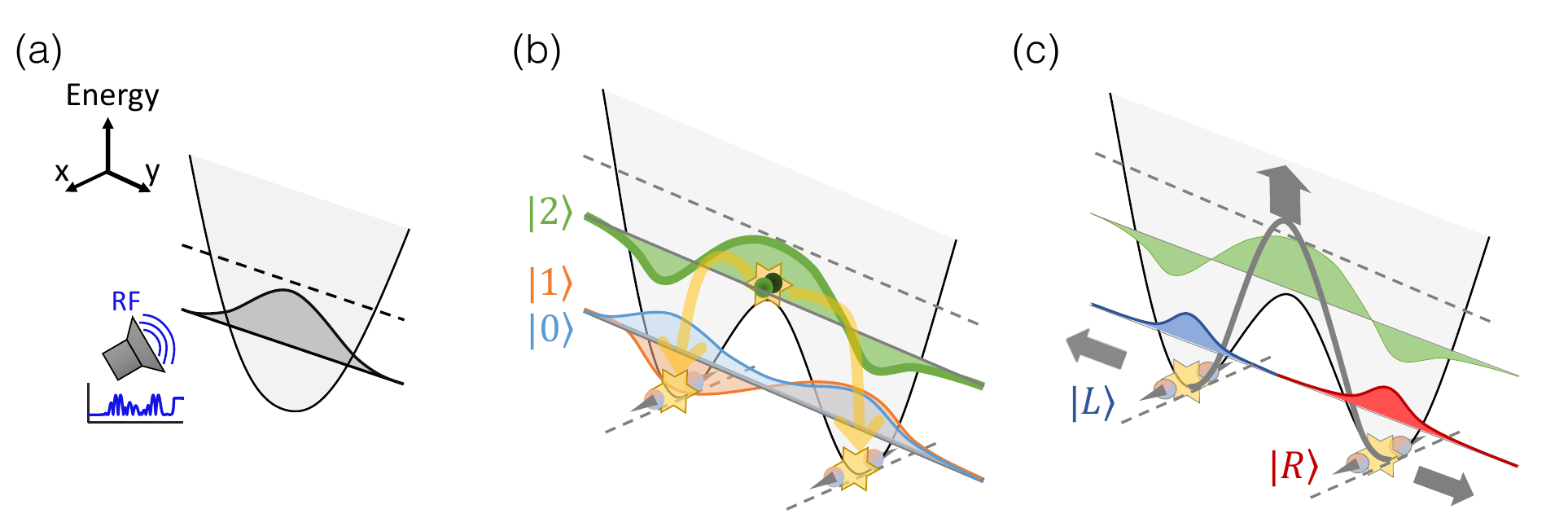}
        \caption{Sketch of the experimental procedure. (a) The quasi-BEC (gray) lies initially in the transverse ground state of a single-well potential characterized by a tightly confined direction (y-axis or transverse axis) and a weakly confined direction (x-axis or longitudinal axis, potential curve along this axis not displayed). An RF-field with variable amplitude is used to excite the condensate and at the same time reach a double-well configuration along the y-axis. (b) The final double-well potential with its vibrational states along the y-axis: the second-excited state (green), which constitutes the \textit{source state}, the first-excited (orange) and the ground state (blue) are defined by $\ket{n_y}$, where $n_y=0,1,2$ is the vibrational quantum number.
        Two atoms from the source state can collide and decay into a twin pair (opposite momenta along the x-axis). Since the atoms in the twin pair can either be emitted in the symmetric $\ket{0}$ (blue) or the anti-symmetric $\ket{1}$ (orange) transverse state, we define the emitted two-particle state as \textit{double twin-atom beam} (DTB) state. (c) The DTB state can also be expressed in terms of the localized left- $\ket{L}$ (blue curve) and right-well state $\ket{R}$ (red curve). The grey arrows represent the process of quickly lifting up the barrier height and pushing the well's minima away from each other.}
        \label{fig:DTBsketch}
    \end{figure*}

Our experiment starts with preparing a one-dimensional (1D) quasi-Bose-Eistein Condensate (BEC)~\cite{petrov2000regimes} of 600-2000 atoms (\(T \lesssim\) \SI{40}{\nano\kelvin}) magnetically trapped in a tight transverse anharmonic potential (\(\nu_{y,z} \simeq \) \SI{2}{\kilo\hertz}) with a shallow longitudinal harmonic confinement (\(\nu_x \simeq \) \SI{10}{\hertz}) created below an atom chip~\cite{FolmanPRL2000}. 
The experimental procedure to create the DTBs [Fig.~\ref{fig:DTBsketch}(a) and \ref{fig:DTBsketch}(b)] begins with splitting the 1D trapping potential into a double-well potential~\cite{Schumm2005b}. The splitting ramp is designed by optimal control to achieve \textit{state inversion}, that is, the 1D quasi-BEC is transferred to the second transversely excited state of the double-well potential, the desired \textit{source state}. 
In particular, a potential barrier with a time-varying height is first created along the $y$ axis. During a certain time interval, the barrier height is lifted up and down symmetrically with respect to the minima of the two wells and finally settled at a value which determines the final double-well geometry. When the amplitude of the RF-field is increased, the distance between the two minima increases.
The manipulation of the transverse potential is achieved by radio-frequency dressing~\cite{hofferberth2006radiofrequency,LesanovskyPRA2006}. The precise amplitude of the applied rf field is determined by optimal-control techniques (see Supplemental Material~\cite{SM}).

The final potential along the $y$ axis is displayed in Fig.~\ref{fig:DTBsketch}(b), together with the corresponding single-particle eigenstates. The states are labelled $\ket{n_y}$ with the vibrational quantum number along the y-axis $n_y = 0,1,2$ (since along the other transverse direction $n_z=0$ during the whole experiment, we have dropped the corresponding index). The second excited state (green) $\ket{2}$ has an energy \(\epsilon/h = \nu_2-\nu_0 \simeq \) \SI{1.3}{\kilo\hertz} and represents the \textit{source state}. The two lowest eigenstates, $\ket{0}$ (light blue) and $\ket{1}$ (orange), have an energy difference $E_1-E_0 \ll \text{min} \{ \epsilon, \mu \}$, where $\mu$ is the chemical potential~\cite{gerbier2004quasi}, and thus are assumed to be degenerate. In Fig.~\ref{fig:DTBsketch}(c), the localized left- $\ket{L}$ and right-well state $\ket{R}$ of the double-well potential are displayed (blue and red curves, respectively). The two basis representations are linked by the relations $\ket{0} = (\ket{L}+\ket{R})/\sqrt{2}$ and $\ket{1} = (\ket{L}-\ket{R})/\sqrt{2}$.

A binary collision between two atoms in the source state can lead to the emission of a pair of twin atoms (for an extensive study of the emission process see Ref.~\cite{lewis2020atomic}). Because of momentum conservation, the atoms are emitted with opposite momenta along the shallow longitudinal direction ($x$ axis), which constitutes the first pair of modes available to each indistinguishable atom. 
The residual potential energy $\epsilon$ from the source state gets translated into kinetic energy of the emitted twin pairs. This determines a selection of only two longitudinal momenta $\pm k_0=\pm \sqrt{2m\epsilon}/\hbar$.
Furthermore, the presence of a double-well potential along the tightly-confined transverse direction ($y$ axis) defines an additional spatial degree of freedom represented by the left $\ket{L}$ and right state $\ket{R}$ in Fig.~\ref{fig:DTBsketch}(c), thus bringing to four the total number of modes available to each indistinguishable atom.

The twin pair is created by $s$-wave scattering ($\delta$-function interaction) between two bosonic atoms in the source state and emitted along the symmetric double waveguide with negligible overlap between the $\ket{L}$ and $\ket{R}$ states. For bosonic particles the state of the atom pair is expected to be in the maximally entangled state:

\begin{equation}\label{eq:Bell-state}
    \ket{\Psi_{DTB}} = \frac{1}{\sqrt{2}}(\ket{L}_-\ket{L}_+ + \ket{R}_-\ket{R}_+),
\end{equation}
where $\ket{i}_-\ket{i}_+ \equiv \ket{i}_{-k_0} \otimes \ket{i}_{+k_0}$ and $i=\{L,R\}$ (for details on the calculation leading to this result see the Supplemental Material~\cite{SM}).
Such a two-particle state is hereafter denoted as DTB state.

\begin{figure*}[htb]
        \includegraphics[width=0.8\textwidth]{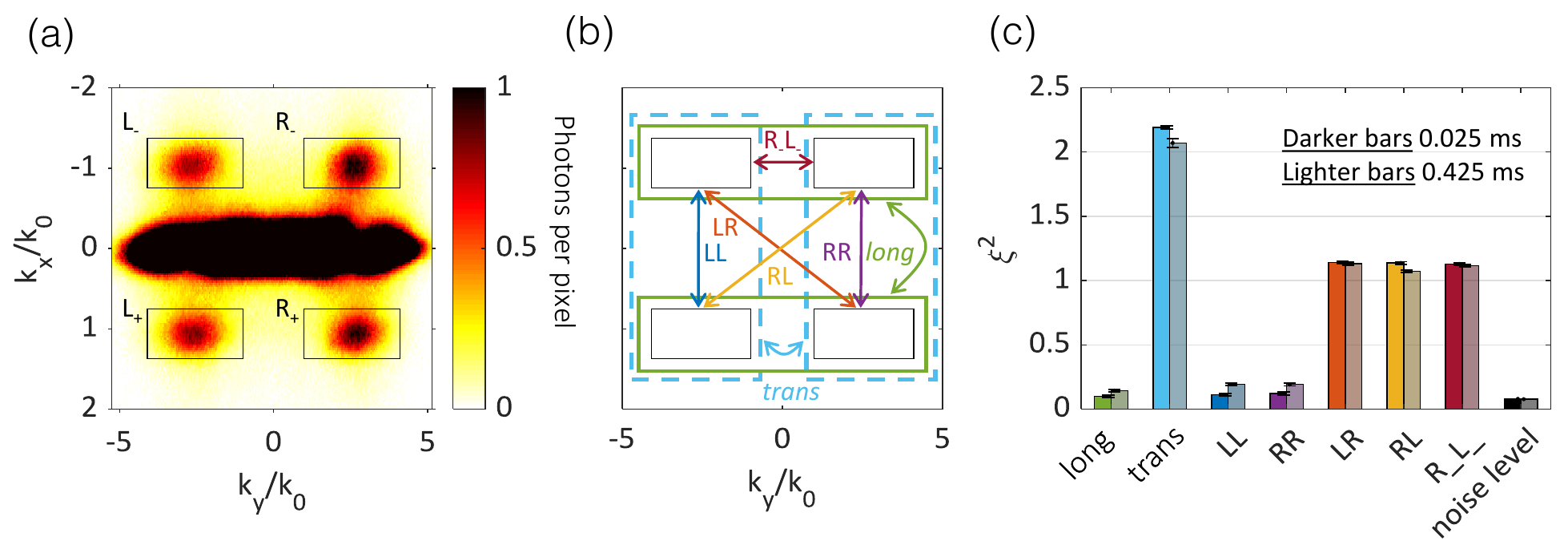}
        \caption{Intra-mode correlations. (a) Experimental fluorescence image averaged over 825 experimental runs obtained with the separation procedure. Each run involves 2000-2200 total atoms, in average 150 of which are DTB atoms (75 pairs). The long time-of-flight makes the initial y-axis momentum distribution accessible (see Supplemental Material~\cite{SM}). The central cloud corresponds to the source state, while the emitted DTB atoms are found at $\pm k_0$. The black boxes define the regions used for the correlation analysis. (b) Color-scheme definition of the different combinations of DTB modes considered. (c) Histogram of different number-squeezing values $\xi^2$ for each combination of DTB modes defined in (b). 
        \label{fig:figure2}}
    \end{figure*}   
Experimental evidence will be provided hereafter in favour of the generation of the state in Eq.~(\ref{eq:Bell-state}). First, in the so-called \textit{separation procedure} we will measure the classical correlations among the different four single-particle modes. To do so, we quickly increase the potential barrier separating the two waveguides before the trap is switched off. This imparts a large and opposite transverse momentum onto the L- and R-well states, so that they can be counted separately. The correlation analysis then lets us exclude the emission of $\ket{L}_-\ket{R}_+$ and $\ket{R}_-\ket{L}_+$ pairs. However, the same analysis cannot exclude the presence of mixed states of $\ket{L}_-\ket{L}_+$ and $\ket{R}_-\ket{R}_+$ with no coherent superposition. 
Therefore, in the so-called \textit{interference procedure} we release the atomic wavefunctions of the emitted beams from the two waveguides; they transversally expand, overlap and interfere. A second-order correlation analysis will then reveal coherent superposition between a pair being emitted into the L waveguide and the same pair being emitted into the R waveguide, hence excluding the presence of only mixed states of such twin pairs. Moreover, the specific quantum superposition detected in this experiment is consistent with the predicted zero relative phase between the L and R twin pairs in Eq.~(\ref{eq:Bell-state}).

Independently of the experimental procedure, the trap is held for a certain holding time $t_{hold}$. The BEC undergoes a free-fall stage and expands for a time of flight of \SI{44}{\ms} before the atoms are detected by traversing the light sheet of our single atom imaging detector~\cite{bucker2009single}. Because of the long time of flight, the image shows the $y$ axis \textit{in situ} momentum distribution of the atoms (see Supplemental Material~\cite{SM}).

\textit{Separation procedure.\textemdash}In order to resolve the transverse states, we imprint an extra transverse acceleration. This is done by a quick rise of the potential barrier between the L- and R-well [Fig.~\ref{fig:DTBsketch}(c)]. 

A typical image resulting from the separation procedure, averaged over many repetitions, is plotted in Fig.~\ref{fig:figure2}(a). This set of data involves an average of $75$ DTB pairs produced in each repetition. The averaged image shows the remaining BEC at the center and four DTB zones: $L_-$, $R_-$, $L_+$, $R_+$ (black boxes), defined by the two transverse states $\ket{L}$ and $\ket{R}$ and the two longitudinal momenta $\pm k_0$.
We consider the correlations among two signals contained in any pair of the black boxes defined in Fig.~\ref{fig:figure2}(a). This defines a certain number of combinations of two DTB modes, each of which is labeled with an index [see Fig.~\ref{fig:figure2}(b)]. For each combination of modes, we compute the value of the number-squeezing parameter:
\be
\xi^2 = \frac{\Delta S_-^2 }{ \Delta_b S_-^2}-\xi^2_n, 
\ee
where $\Delta S_-^2$ represents the variance of the signal difference $S_-$ between the two boxes considered, $\Delta_b S_-^2$ denotes the corresponding binomial variance and $\xi^2_n$ the noise contribution to the squeezing parameter (see the Supplemental Material~\cite{SM}). A value of $\xi^2<1$ defines a number-squeezed emission.

In Fig.~\ref{fig:figure2}(c), the number-squeezing value $\xi^2$ is displayed as vertical bars as a function of the different combinations of DTB modes considered (the actual values are also expressed in Table \ref{tab:table3}). 
We observe that the different combinations of DTB modes can be classified in three groups depending on the value of the number squeezing: (a) $\xi^2 \approx 0$ for $LL, RR, long$ (b) $\xi^2 \approx 1$ for $LR, RL, R_-L_-$ (c) $\xi^2 \approx 2$ for $trans$. The group (a) refers to correlations between atoms that have opposite longitudinal momenta and belong to the same waveguide ($LL$ or $RR$) or to any of them ($long$). This characteristic defines atoms belonging to the same twin pair [see Eq.~(\ref{eq:Bell-state})]; hence we find $\xi^2 < 1$. The group (b) refers to atoms that do \textit{not} belong to the same twin pair, either because these combinations of DTB modes mix different waveguides ($LR$ and $RL$) or because they consider atoms with the same longitudinal momenta ($R_-L_-$); hence the signals are uncorrelated and we find $\xi^2 \approx 1$. The last group (c) contains the combination $trans$, which compares the \textit{total} signal between the $L$ and $R$ waveguides.
Given the state in Eq.~(\ref{eq:Bell-state}), we expect twin \textit{pairs} to be detected either in the $L$ or in the $R$ waveguide, without correlation between these two modes. However, each atom is part of a twin pair, so the \textit{atom} detection is not $trans$ uncorrelated. In terms of the statistics of individual atoms, we find $\xi^2_{trans}=2$ (see also the Supplemental Material~\cite{SM}).

\begin{table}
    \centering
    \caption{Number squeezing. Noise-corrected atom number squeezing for different combinations of the four DTB modes.}
    \begin{tabular}{ c c c c c c c c c }
        \toprule
        $t_{hold} (\SI{}{\ms})$  & $\xi^2_{LL}$ & $\xi^2_{RR}$ &
              $\xi^2_{LR}$ & $\xi^2_{RL}$ & $\xi^2_{L_-R_-}$ & $\xi^2_{L/R}$ & $\xi^2_{k_0/\text{-}k_0}$ & $\xi^2_n$\\
              \hline
        0.025 &  0.11 & 0.12 & 1.14 & 1.14 & 1.13 & 2.19 & 0.10 & 0.078\\
        0.425 & 0.19 & 0.19 & 1.13 & 1.07 & 1.12 & 2.07 & 0.14 & 0.076\\
        \bottomrule
    \end{tabular}
    
    \label{tab:table3}
\end{table}

\begin{figure*}
    \includegraphics[width=0.8\textwidth]{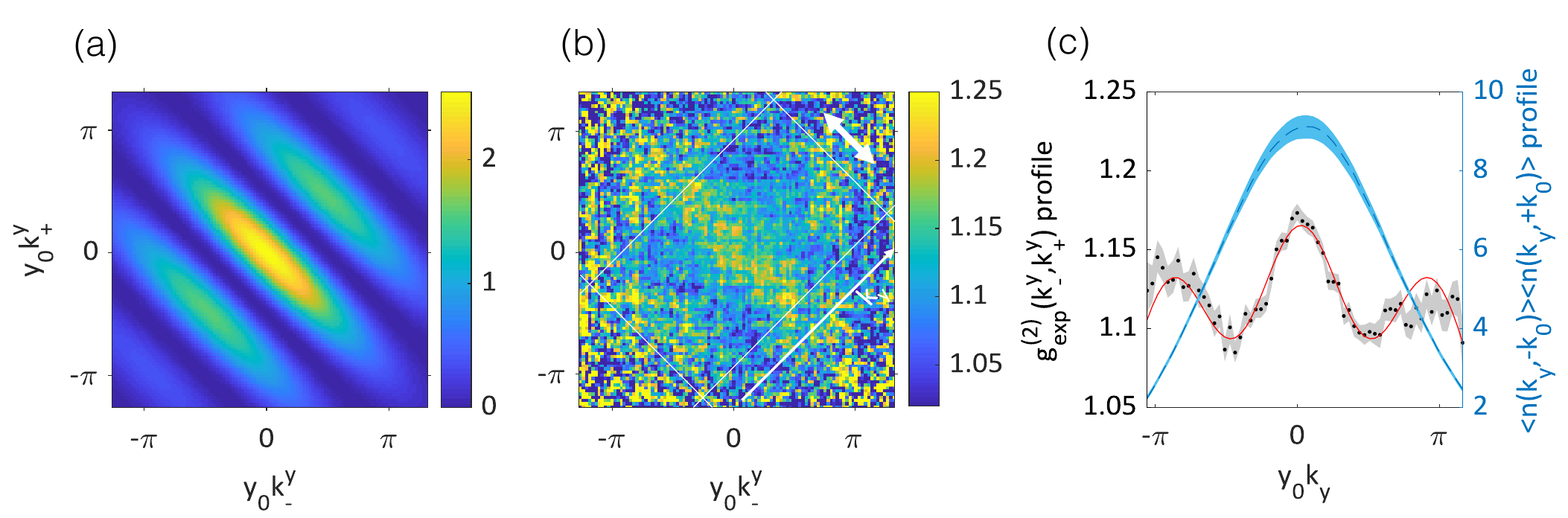}
    \caption{Two-particle interference pattern. 
    (a) Theoretical un-normalized $G^{(2)}(k^y_-,k^y_+) = \langle n(k_y,-k_0) n(k_y,+k_0) \rangle$ pattern assuming a DTB state of the form Eq.~\ref{eq:Bell-state}. (b) Experimental $g^{(2)}_{exp}(k^y_-,k^y_+)$ pattern. This set of data involves 1498 experimental runs, where each run contains 700-760 total atoms, 20 of which are DTB atoms (10 pairs), on average. (c) One-dimensional mean profiles obtained from averaging along the diagonal within the white box superimposed in Fig.~\ref{fig:g2fringe}b. 
    The mean anti-diagonal profile of $g^{(2)}_{exp}(k^y_-,k^y_+)$ (black dots) is compared to the mean anti-diagonal profile of $\langle n(k_y,-k_0)\rangle \langle n(k_y,+k_0) \rangle$ (light blue curve). The red curve represents a fit of the data from which we extract a value of the contrast of the two-atom interference $C= 0.032 \pm  0.004$. Units are scaled by the diagonal $\sqrt{2}$ factor and then normalized by the wells spacing $2 y_0 = 1.3~\mu m $. The shaded areas represent the standard error of the mean. 
    \label{fig:g2fringe}}
\end{figure*}

These results are compatible with the generation of a maximally entangled state as in Eq.~(\ref{eq:Bell-state}), but also with a two-particle mixed state of $\ket{L}_-\ket{L}_+$ and $\ket{R}_-\ket{R}_+$. To exclude this case we need to look at the two-particle interference pattern.

\textit{Interference procedure.\textemdash} 
In our experiment, each twin pair can be emitted in either the $L$ or $R$ waveguide. These represent two two-particle quantum paths that interfere with equal amplitude (balanced double-well) when performing an interference measurement procedure; i.e., we avoid imprinting an extra transverse acceleration [Fig.~\ref{fig:DTBsketch}(c)]. 
Unlike the single-particle case where an interference pattern is visible already in the mean density in momentum space (one-particle property), in the two-particle case we need to look at two-particle properties in order to extract information on the final state~\cite{bonneau2018characterizing}.


If the DTB emission preserves the coherence of the quasi-BEC, the DTB state shows two-atom interference in the second-order correlation function $g^{(2)}(k^y_-,k^y_+)$ linking atoms of opposite momenta:
    \begin{equation}\label{eq:g2}
        g^{(2)}(k^y_-,k^y_+) = \frac{\langle n(k_y,-k_0) n(k_y,+k_0) \rangle}{\langle n(k_y,-k_0)\rangle \langle n(k_y,+k_0)\rangle},
    \end{equation}
where $k_y$ is the transverse wave-vector and $n(k_y,\pm k_0)$ is the single-particle density profile along the transverse axis at the two longitudinal momenta $\pm k_0$.
The particular fringe pattern in $g^{(2)}(k^y_-,k^y_+)$ depends on the underlying density matrix associated to the DTB state~\cite{bonneau2018characterizing}.
Maximal contrast requires identifying the partners in each atom pair. 
In a low-pair emission regime, we emit an average of $10$ DTB pairs in each experimental run. Averaging over the pairs will reduce the contrast in the observed interference. 

In Figs.~\ref{fig:g2fringe}(a) and \ref{fig:g2fringe}(b), we compare the simulated unnormalized $G^{(2)}(k^y_-,k^y_+) = \langle n(k_y,-k_0) n(k_y,+k_0) \rangle$ and experimental $g^{(2)}_{exp}(k^y_-,k^y_+)$ patterns: Fig.~\ref{fig:g2fringe}(a) shows the theoretical fringe pattern assuming a two-particle state of the form Eq.~(\ref{eq:Bell-state}); Fig.~\ref{fig:g2fringe}(b) shows the experimental $g^{(2)}_{exp}(k^y_-,k^y_+)$ pattern averaged over 1498 experimental runs. The number of visible fringes depends on the value of the wells spacing $2 y_0$ between the two potential waveguides. In order to compare the theoretical pattern (a) with the experimental one (b), we use $2 y_0=1.3~\mu m$. This value is obtained from a simulation of the final double-well potential that was calibrated to match with the experiment. 

The white box in Fig.~\ref{fig:g2fringe}(b), defines the integration area for the profiles in Fig.~\ref{fig:g2fringe}(c): the double-arrow defines the integration axis, while the single arrow illustrates the transverse momentum coordinate $k_y$ [horizontal axis in Fig.~\ref{fig:g2fringe}(c)]. The projected pattern shows clear fringes with a period consistent with the double well and a contrast $C = 0.032 \pm  0.004$. In order to ensure that the central fringe is not originating from the envelope, we compare the fringe profile with the the mean profile obtained considering only the product of the independently averaged profiles $\langle n(k_y,-k_0)\rangle \langle n(k_y,+k_0)\rangle$ (blue dashed curve). 

This fringe pattern in the measured $g^{(2)}(k^y_-,k^y_+)$ [Fig.~\ref{fig:g2fringe}(b) and \ref{fig:g2fringe}(c)] combined with the absence of an interference fringe in the single-particle density is one of the central results of our experiment and it constitutes direct evidence for genuine two-particle interference. For a statistical mixture of the states $\ket{L}_-\ket{L}_+$ and $\ket{R}_-\ket{R}_+$, one would expect a flat profile $g^{(2)}(k^y_-,k^y_+)=1$.  
Combined with the measurements of the number-squeezing correlations between the four guided DTB modes in Table \ref{tab:table3} and following Ref.~\cite{bonneau2018characterizing}, our experiment shows that a significant fraction of atom pairs are emitted in the maximally entangled state of Eq.~(\ref{eq:Bell-state}). This is a ``lucky'' situation where the reconstruction of the full density matrix of the two-particle state (and hence an entanglement demonstration) is in principle possible without any phase rotation, just by looking at the two-particle interference pattern~\cite{bonneau2018characterizing}.
We attribute the low contrast of $C = 0.032 \pm  0.004$ in our present experiment to the relatively large number of 10 pairs emitted on average in each measurement, thereby washing out the interference pattern.

Our experiments show a path towards a quantitative demonstration of entanglement for propagating atom beams in such a system as suggested in~\cite{bonneau2018characterizing}. In order to achieve this goal we need to significantly increase the contrast of the two-particle interference, which will require a more detailed study of the emission process and better control over the number of emitted pairs, down to experiment with single pairs. A phase shift can be applied to the propagating DTBs by tilting the double-well potential to introduce an energy difference between the left- and right-well states, as in~\cite{pigneur2018relaxation}. As an alternative procedure, one could implement Bragg deflectors as in~\cite{lewis2015proposal,dussarrat2017two} to rotate the state after its generation.
 
As a more general outlook, we see a huge potential in exploring non-linear matter-wave optics for atoms propagating in waveguides and integrated matter-wave circuits. The processes behind the twin-atom emission are closely related to the matter-wave equivalents of parametric amplification and four-wave mixing. We envision the development of non-linear matter-wave quantum optics. The creation of entangled atom-laser beams in twin-beam emission above threshold would be one directly accessible example. 

\section*{Acknowledgements} \label{sec:acknowledgements}

        We thank M. Pigneur and R. Trubko for help in the initial phases of the project, K. Nemoto and W.J. Munro for numerous inspiring discussions and C. L\'{e}v\^{e}que for help with the manuscript. 
        This work was funded by the Austrian Science Fund (FWF) via the QUANTERA project CEBBEC (I 3759-N27) and  
        M.M. and P.H. acknowledge support by the project LATIN (Y-1121). I.M. acknowledges the support by the Wiener  Wissenschafts- und Technologiefonds (WWTF) via Grant No. MA16-066 (SEQUEX) 
        and by the Austrian Science Fund (FWF) via Grant SFB F65 (Complexity in PDE systems).
        T.Z. is supported by the EU's Horizon 2020 program under the  M. Curie Grant Nos. 765267 (QuSCo). F.B., M.M. and T.Z. acknowledge the support from CoQuS.
        M.B. was supported by the EU through the M. Curie Grant ETAB (MSCA-IF-2014-EF 656530) and by the Austrian Science Fund (FWF) through the Lise Meitner Grant CoPaNeq (M2088-M27). S.M. acknowledges support from the Italian PRIN 2017.
        T.C. and S.M. acknowledge the support of the EC Quantum Flagship project PASQUANS.



\begin{thebibliography}{10}%
\makeatletter
\providecommand \@ifxundefined [1]{%
 \@ifx{#1\undefined}
}%
\providecommand \@ifnum [1]{%
 \ifnum #1\expandafter \@firstoftwo
 \else \expandafter \@secondoftwo
 \fi
}%
\providecommand \@ifx [1]{%
 \ifx #1\expandafter \@firstoftwo
 \else \expandafter \@secondoftwo
 \fi
}%
\providecommand \natexlab [1]{#1}%
\providecommand \enquote  [1]{``#1''}%
\providecommand \bibnamefont  [1]{#1}%
\providecommand \bibfnamefont [1]{#1}%
\providecommand \citenamefont [1]{#1}%
\providecommand \href@noop [0]{\@secondoftwo}%
\providecommand \href [0]{\begingroup \@sanitize@url \@href}%
\providecommand \@href[1]{\@@startlink{#1}\@@href}%
\providecommand \@@href[1]{\endgroup#1\@@endlink}%
\providecommand \@sanitize@url [0]{\catcode `\\12\catcode `\$12\catcode
  `\&12\catcode `\#12\catcode `\^12\catcode `\_12\catcode `\%12\relax}%
\providecommand \@@startlink[1]{}%
\providecommand \@@endlink[0]{}%
\providecommand \url  [0]{\begingroup\@sanitize@url \@url }%
\providecommand \@url [1]{\endgroup\@href {#1}{\urlprefix }}%
\providecommand \urlprefix  [0]{URL }%
\providecommand \Eprint [0]{\href }%
\providecommand \doibase [0]{http://dx.doi.org/}%
\providecommand \selectlanguage [0]{\@gobble}%
\providecommand \bibinfo  [0]{\@secondoftwo}%
\providecommand \bibfield  [0]{\@secondoftwo}%
\providecommand \translation [1]{[#1]}%
\providecommand \BibitemOpen [0]{}%
\providecommand \bibitemStop [0]{}%
\providecommand \bibitemNoStop [0]{.\EOS\space}%
\providecommand \EOS [0]{\spacefactor3000\relax}%
\providecommand \BibitemShut  [1]{\csname bibitem#1\endcsname}%
\let\auto@bib@innerbib\@empty
\bibitem [{\citenamefont {van Frank}\ \emph {et~al.}(2014)\citenamefont {van
  Frank}, \citenamefont {Negretti}, \citenamefont {Berrada}, \citenamefont
  {B{\"u}cker}, \citenamefont {Montangero}, \citenamefont {Schaff},
  \citenamefont {Schumm}, \citenamefont {Calarco},\ and\ \citenamefont
  {Schmiedmayer}}]{FrankNC2014}%
  \BibitemOpen
  \bibfield  {author} {\bibinfo {author} {\bibfnamefont {S.}~\bibnamefont {van
  Frank}}, \bibinfo {author} {\bibfnamefont {A.}~\bibnamefont {Negretti}},
  \bibinfo {author} {\bibfnamefont {T.}~\bibnamefont {Berrada}}, \bibinfo
  {author} {\bibfnamefont {R.}~\bibnamefont {B{\"u}cker}}, \bibinfo {author}
  {\bibfnamefont {S.}~\bibnamefont {Montangero}}, \bibinfo {author}
  {\bibfnamefont {J.-F.}\ \bibnamefont {Schaff}}, \bibinfo {author}
  {\bibfnamefont {T.}~\bibnamefont {Schumm}}, \bibinfo {author} {\bibfnamefont
  {T.}~\bibnamefont {Calarco}}, \ and\ \bibinfo {author} {\bibfnamefont
  {J.}~\bibnamefont {Schmiedmayer}},\ }\href@noop {} {\bibfield  {journal}
  {\bibinfo  {journal} {Nature communications}\ }\textbf {\bibinfo {volume}
  {5}},\ \bibinfo {pages} {1} (\bibinfo {year} {2014})}\BibitemShut {NoStop}%
\bibitem [{\citenamefont {van Frank}\ \emph {et~al.}(2016)\citenamefont {van
  Frank}, \citenamefont {Bonneau}, \citenamefont {Schmiedmayer}, \citenamefont
  {Hild}, \citenamefont {Gross}, \citenamefont {Cheneau}, \citenamefont
  {Bloch}, \citenamefont {Pichler}, \citenamefont {Negretti}, \citenamefont
  {Calarco} \emph {et~al.}}]{FrankSR2016}%
  \BibitemOpen
  \bibfield  {author} {\bibinfo {author} {\bibfnamefont {S.}~\bibnamefont {van
  Frank}}, \bibinfo {author} {\bibfnamefont {M.}~\bibnamefont {Bonneau}},
  \bibinfo {author} {\bibfnamefont {J.}~\bibnamefont {Schmiedmayer}}, \bibinfo
  {author} {\bibfnamefont {S.}~\bibnamefont {Hild}}, \bibinfo {author}
  {\bibfnamefont {C.}~\bibnamefont {Gross}}, \bibinfo {author} {\bibfnamefont
  {M.}~\bibnamefont {Cheneau}}, \bibinfo {author} {\bibfnamefont
  {I.}~\bibnamefont {Bloch}}, \bibinfo {author} {\bibfnamefont
  {T.}~\bibnamefont {Pichler}}, \bibinfo {author} {\bibfnamefont
  {A.}~\bibnamefont {Negretti}}, \bibinfo {author} {\bibfnamefont
  {T.}~\bibnamefont {Calarco}},  \emph {et~al.},\ }\href@noop {} {\bibfield
  {journal} {\bibinfo  {journal} {Scientific reports}\ }\textbf {\bibinfo
  {volume} {6}},\ \bibinfo {pages} {34187} (\bibinfo {year}
  {2016})}\BibitemShut {NoStop}%
\bibitem [{\citenamefont {Lesanovsky}\ \emph {et~al.}(2006)\citenamefont
  {Lesanovsky}, \citenamefont {Schumm}, \citenamefont {Hofferberth},
  \citenamefont {Andersson}, \citenamefont {Kr{\"u}ger},\ and\ \citenamefont
  {Schmiedmayer}}]{LesanovskyPRA2006}%
  \BibitemOpen
  \bibfield  {author} {\bibinfo {author} {\bibfnamefont {I.}~\bibnamefont
  {Lesanovsky}}, \bibinfo {author} {\bibfnamefont {T.}~\bibnamefont {Schumm}},
  \bibinfo {author} {\bibfnamefont {S.}~\bibnamefont {Hofferberth}}, \bibinfo
  {author} {\bibfnamefont {L.~M.}\ \bibnamefont {Andersson}}, \bibinfo {author}
  {\bibfnamefont {P.}~\bibnamefont {Kr{\"u}ger}}, \ and\ \bibinfo {author}
  {\bibfnamefont {J.}~\bibnamefont {Schmiedmayer}},\ }\href@noop {} {\bibfield
  {journal} {\bibinfo  {journal} {Physical Review A}\ }\textbf {\bibinfo
  {volume} {73}},\ \bibinfo {pages} {033619} (\bibinfo {year}
  {2006})}\BibitemShut {NoStop}%
\bibitem [{\citenamefont {Caneva}\ \emph {et~al.}(2011)\citenamefont {Caneva},
  \citenamefont {Calarco},\ and\ \citenamefont {Montangero}}]{CanevaPRA2011}%
  \BibitemOpen
  \bibfield  {author} {\bibinfo {author} {\bibfnamefont {T.}~\bibnamefont
  {Caneva}}, \bibinfo {author} {\bibfnamefont {T.}~\bibnamefont {Calarco}}, \
  and\ \bibinfo {author} {\bibfnamefont {S.}~\bibnamefont {Montangero}},\
  }\href {\doibase 10.1103/PhysRevA.84.022326} {\bibfield  {journal} {\bibinfo
  {journal} {Phys. Rev. A}\ }\textbf {\bibinfo {volume} {84}},\ \bibinfo
  {pages} {022326} (\bibinfo {year} {2011})}\BibitemShut {NoStop}%
\bibitem [{\citenamefont {Zhao}\ \emph {et~al.}(2007)\citenamefont {Zhao},
  \citenamefont {Chen}, \citenamefont {Pan}, \citenamefont {Schmiedmayer},
  \citenamefont {Recati}, \citenamefont {Astrakharchik},\ and\ \citenamefont
  {Calarco}}]{zhao2007high}%
  \BibitemOpen
  \bibfield  {author} {\bibinfo {author} {\bibfnamefont {B.}~\bibnamefont
  {Zhao}}, \bibinfo {author} {\bibfnamefont {Z.-B.}\ \bibnamefont {Chen}},
  \bibinfo {author} {\bibfnamefont {J.-W.}\ \bibnamefont {Pan}}, \bibinfo
  {author} {\bibfnamefont {J.}~\bibnamefont {Schmiedmayer}}, \bibinfo {author}
  {\bibfnamefont {A.}~\bibnamefont {Recati}}, \bibinfo {author} {\bibfnamefont
  {G.~E.}\ \bibnamefont {Astrakharchik}}, \ and\ \bibinfo {author}
  {\bibfnamefont {T.}~\bibnamefont {Calarco}},\ }\href@noop {} {\bibfield
  {journal} {\bibinfo  {journal} {Physical Review A}\ }\textbf {\bibinfo
  {volume} {75}},\ \bibinfo {pages} {042312} (\bibinfo {year}
  {2007})}\BibitemShut {NoStop}%
\bibitem [{\citenamefont {B{\"u}cker}\ \emph {et~al.}(2011)\citenamefont
  {B{\"u}cker}, \citenamefont {Grond}, \citenamefont {Manz}, \citenamefont
  {Berrada}, \citenamefont {Betz}, \citenamefont {Koller}, \citenamefont
  {Hohenester}, \citenamefont {Schumm}, \citenamefont {Perrin},\ and\
  \citenamefont {Schmiedmayer}}]{bucker2011twin}%
  \BibitemOpen
  \bibfield  {author} {\bibinfo {author} {\bibfnamefont {R.}~\bibnamefont
  {B{\"u}cker}}, \bibinfo {author} {\bibfnamefont {J.}~\bibnamefont {Grond}},
  \bibinfo {author} {\bibfnamefont {S.}~\bibnamefont {Manz}}, \bibinfo {author}
  {\bibfnamefont {T.}~\bibnamefont {Berrada}}, \bibinfo {author} {\bibfnamefont
  {T.}~\bibnamefont {Betz}}, \bibinfo {author} {\bibfnamefont {C.}~\bibnamefont
  {Koller}}, \bibinfo {author} {\bibfnamefont {U.}~\bibnamefont {Hohenester}},
  \bibinfo {author} {\bibfnamefont {T.}~\bibnamefont {Schumm}}, \bibinfo
  {author} {\bibfnamefont {A.}~\bibnamefont {Perrin}}, \ and\ \bibinfo {author}
  {\bibfnamefont {J.}~\bibnamefont {Schmiedmayer}},\ }\href@noop {} {\bibfield
  {journal} {\bibinfo  {journal} {Nature Physics}\ }\textbf {\bibinfo {volume}
  {7}},\ \bibinfo {pages} {608} (\bibinfo {year} {2011})}\BibitemShut {NoStop}%
\bibitem [{\citenamefont {B{\"u}cker}\ \emph {et~al.}(2009)\citenamefont
  {B{\"u}cker}, \citenamefont {Perrin}, \citenamefont {Manz}, \citenamefont
  {Betz}, \citenamefont {Koller}, \citenamefont {Plisson}, \citenamefont
  {Rottmann}, \citenamefont {Schumm},\ and\ \citenamefont
  {Schmiedmayer}}]{bucker2009single}%
  \BibitemOpen
  \bibfield  {author} {\bibinfo {author} {\bibfnamefont {R.}~\bibnamefont
  {B{\"u}cker}}, \bibinfo {author} {\bibfnamefont {A.}~\bibnamefont {Perrin}},
  \bibinfo {author} {\bibfnamefont {S.}~\bibnamefont {Manz}}, \bibinfo {author}
  {\bibfnamefont {T.}~\bibnamefont {Betz}}, \bibinfo {author} {\bibfnamefont
  {C.}~\bibnamefont {Koller}}, \bibinfo {author} {\bibfnamefont
  {T.}~\bibnamefont {Plisson}}, \bibinfo {author} {\bibfnamefont
  {J.}~\bibnamefont {Rottmann}}, \bibinfo {author} {\bibfnamefont
  {T.}~\bibnamefont {Schumm}}, \ and\ \bibinfo {author} {\bibfnamefont
  {J.}~\bibnamefont {Schmiedmayer}},\ }\href@noop {} {\bibfield  {journal}
  {\bibinfo  {journal} {New Journal of Physics}\ }\textbf {\bibinfo {volume}
  {11}},\ \bibinfo {pages} {103039} (\bibinfo {year} {2009})}\BibitemShut
  {NoStop}%
\bibitem [{\citenamefont {Pigneur}(2020)}]{pigneurnon}%
  \BibitemOpen
  \bibfield  {author} {\bibinfo {author} {\bibfnamefont {M.}~\bibnamefont
  {Pigneur}},\ }\href@noop {} {\emph {\bibinfo {title} {Non-equilibrium
  Dynamics of Tunnel-Coupled Superfluids: Relaxation to a Phase-Locked
  Equilibrium State in a One-Dimensional Bosonic Josephson Junction}}}\
  (\bibinfo  {publisher} {Springer Nature},\ \bibinfo {year}
  {2020})\BibitemShut {NoStop}%
\bibitem [{\citenamefont {B{\"u}cker}(2013)}]{bucker2013twin}%
  \BibitemOpen
  \bibfield  {author} {\bibinfo {author} {\bibfnamefont {R.}~\bibnamefont
  {B{\"u}cker}},\ }\emph {\bibinfo {title} {Twin-atom beam generation in a
  one-dimensional Bose gas}},\ \href@noop {} {Ph.D. thesis},\ \bibinfo
  {school} {Techische Universit\"at Wien} (\bibinfo {year} {2013})\BibitemShut
  {NoStop}%
\bibitem [{\citenamefont {Basden}\ \emph {et~al.}(2003)\citenamefont {Basden},
  \citenamefont {Haniff},\ and\ \citenamefont {Mackay}}]{basden2003photon}%
  \BibitemOpen
  \bibfield  {author} {\bibinfo {author} {\bibfnamefont {A.}~\bibnamefont
  {Basden}}, \bibinfo {author} {\bibfnamefont {C.}~\bibnamefont {Haniff}}, \
  and\ \bibinfo {author} {\bibfnamefont {C.}~\bibnamefont {Mackay}},\
  }\href@noop {} {\bibfield  {journal} {\bibinfo  {journal} {Monthly notices of
  the royal astronomical society}\ }\textbf {\bibinfo {volume} {345}},\
  \bibinfo {pages} {985} (\bibinfo {year} {2003})}\BibitemShut {NoStop}%
\end{thebibliography}%


\begin{thebibliography}{34}%
\makeatletter
\providecommand \@ifxundefined [1]{%
 \@ifx{#1\undefined}
}%
\providecommand \@ifnum [1]{%
 \ifnum #1\expandafter \@firstoftwo
 \else \expandafter \@secondoftwo
 \fi
}%
\providecommand \@ifx [1]{%
 \ifx #1\expandafter \@firstoftwo
 \else \expandafter \@secondoftwo
 \fi
}%
\providecommand \natexlab [1]{#1}%
\providecommand \enquote  [1]{``#1''}%
\providecommand \bibnamefont  [1]{#1}%
\providecommand \bibfnamefont [1]{#1}%
\providecommand \citenamefont [1]{#1}%
\providecommand \href@noop [0]{\@secondoftwo}%
\providecommand \href [0]{\begingroup \@sanitize@url \@href}%
\providecommand \@href[1]{\@@startlink{#1}\@@href}%
\providecommand \@@href[1]{\endgroup#1\@@endlink}%
\providecommand \@sanitize@url [0]{\catcode `\\12\catcode `\$12\catcode
  `\&12\catcode `\#12\catcode `\^12\catcode `\_12\catcode `\%12\relax}%
\providecommand \@@startlink[1]{}%
\providecommand \@@endlink[0]{}%
\providecommand \url  [0]{\begingroup\@sanitize@url \@url }%
\providecommand \@url [1]{\endgroup\@href {#1}{\urlprefix }}%
\providecommand \urlprefix  [0]{URL }%
\providecommand \Eprint [0]{\href }%
\providecommand \doibase [0]{http://dx.doi.org/}%
\providecommand \selectlanguage [0]{\@gobble}%
\providecommand \bibinfo  [0]{\@secondoftwo}%
\providecommand \bibfield  [0]{\@secondoftwo}%
\providecommand \translation [1]{[#1]}%
\providecommand \BibitemOpen [0]{}%
\providecommand \bibitemStop [0]{}%
\providecommand \bibitemNoStop [0]{.\EOS\space}%
\providecommand \EOS [0]{\spacefactor3000\relax}%
\providecommand \BibitemShut  [1]{\csname bibitem#1\endcsname}%
\let\auto@bib@innerbib\@empty
\bibitem [{\citenamefont {Dowling}\ and\ \citenamefont
  {Milburn}(2003)}]{dowling2003quantum}%
  \BibitemOpen
  \bibfield  {author} {\bibinfo {author} {\bibfnamefont {J.~P.}\ \bibnamefont
  {Dowling}}\ and\ \bibinfo {author} {\bibfnamefont {G.~J.}\ \bibnamefont
  {Milburn}},\ }\href@noop {} {\bibfield  {journal} {\bibinfo  {journal}
  {Philosophical Transactions of the Royal Society of London. Series A:
  Mathematical, Physical and Engineering Sciences}\ }\textbf {\bibinfo {volume}
  {361}},\ \bibinfo {pages} {1655} (\bibinfo {year} {2003})}\BibitemShut
  {NoStop}%
\bibitem [{Qua(2020)}]{QuantumRoadMap}%
  \BibitemOpen
  \href
  {https://qt.eu//app/uploads/2020/04/Strategic_Research-_Agenda_d_FINAL.pdf}
  {\enquote {\bibinfo {title} {European quantum flagship - strategic research
  agenda},}\ } (\bibinfo {year} {March 2020}),\ \bibinfo {note} {available at
  \url{https://qt.eu//app/uploads/2020/04/Strategic_Research-_Agenda_d_FINAL.pdf}}\BibitemShut
  {NoStop}%
\bibitem [{\citenamefont {Pan}\ \emph {et~al.}(2012)\citenamefont {Pan},
  \citenamefont {Chen}, \citenamefont {Lu}, \citenamefont {Weinfurter},
  \citenamefont {Zeilinger},\ and\ \citenamefont {{\.Z}ukowski}}]{PanRMP2012}%
  \BibitemOpen
  \bibfield  {author} {\bibinfo {author} {\bibfnamefont {J.-W.}\ \bibnamefont
  {Pan}}, \bibinfo {author} {\bibfnamefont {Z.-B.}\ \bibnamefont {Chen}},
  \bibinfo {author} {\bibfnamefont {C.-Y.}\ \bibnamefont {Lu}}, \bibinfo
  {author} {\bibfnamefont {H.}~\bibnamefont {Weinfurter}}, \bibinfo {author}
  {\bibfnamefont {A.}~\bibnamefont {Zeilinger}}, \ and\ \bibinfo {author}
  {\bibfnamefont {M.}~\bibnamefont {{\.Z}ukowski}},\ }\href@noop {} {\bibfield
  {journal} {\bibinfo  {journal} {Reviews of Modern Physics}\ }\textbf
  {\bibinfo {volume} {84}},\ \bibinfo {pages} {777} (\bibinfo {year}
  {2012})}\BibitemShut {NoStop}%
\bibitem [{\citenamefont {Lamehi-Rachti}\ and\ \citenamefont
  {Mittig}(1976)}]{lamehi1976quantum}%
  \BibitemOpen
  \bibfield  {author} {\bibinfo {author} {\bibfnamefont {M.}~\bibnamefont
  {Lamehi-Rachti}}\ and\ \bibinfo {author} {\bibfnamefont {W.}~\bibnamefont
  {Mittig}},\ }\href@noop {} {\bibfield  {journal} {\bibinfo  {journal}
  {Physical Review D}\ }\textbf {\bibinfo {volume} {14}},\ \bibinfo {pages}
  {2543} (\bibinfo {year} {1976})}\BibitemShut {NoStop}%
\bibitem [{\citenamefont {Hensen}\ \emph {et~al.}(2015)\citenamefont {Hensen},
  \citenamefont {Bernien}, \citenamefont {Dr{\'e}au}, \citenamefont {Reiserer},
  \citenamefont {Kalb}, \citenamefont {Blok}, \citenamefont {Ruitenberg},
  \citenamefont {Vermeulen}, \citenamefont {Schouten}, \citenamefont
  {Abell{\'a}n} \emph {et~al.}}]{hensen2015loophole}%
  \BibitemOpen
  \bibfield  {author} {\bibinfo {author} {\bibfnamefont {B.}~\bibnamefont
  {Hensen}}, \bibinfo {author} {\bibfnamefont {H.}~\bibnamefont {Bernien}},
  \bibinfo {author} {\bibfnamefont {A.~E.}\ \bibnamefont {Dr{\'e}au}}, \bibinfo
  {author} {\bibfnamefont {A.}~\bibnamefont {Reiserer}}, \bibinfo {author}
  {\bibfnamefont {N.}~\bibnamefont {Kalb}}, \bibinfo {author} {\bibfnamefont
  {M.~S.}\ \bibnamefont {Blok}}, \bibinfo {author} {\bibfnamefont
  {J.}~\bibnamefont {Ruitenberg}}, \bibinfo {author} {\bibfnamefont {R.~F.}\
  \bibnamefont {Vermeulen}}, \bibinfo {author} {\bibfnamefont {R.~N.}\
  \bibnamefont {Schouten}}, \bibinfo {author} {\bibfnamefont {C.}~\bibnamefont
  {Abell{\'a}n}},  \emph {et~al.},\ }\href@noop {} {\bibfield  {journal}
  {\bibinfo  {journal} {Nature}\ }\textbf {\bibinfo {volume} {526}},\ \bibinfo
  {pages} {682} (\bibinfo {year} {2015})}\BibitemShut {NoStop}%
\bibitem [{\citenamefont {Rowe}\ \emph {et~al.}(2001)\citenamefont {Rowe},
  \citenamefont {Kielpinski}, \citenamefont {Meyer}, \citenamefont {Sackett},
  \citenamefont {Itano}, \citenamefont {Monroe},\ and\ \citenamefont
  {Wineland}}]{rowe2001experimental}%
  \BibitemOpen
  \bibfield  {author} {\bibinfo {author} {\bibfnamefont {M.~A.}\ \bibnamefont
  {Rowe}}, \bibinfo {author} {\bibfnamefont {D.}~\bibnamefont {Kielpinski}},
  \bibinfo {author} {\bibfnamefont {V.}~\bibnamefont {Meyer}}, \bibinfo
  {author} {\bibfnamefont {C.~A.}\ \bibnamefont {Sackett}}, \bibinfo {author}
  {\bibfnamefont {W.~M.}\ \bibnamefont {Itano}}, \bibinfo {author}
  {\bibfnamefont {C.}~\bibnamefont {Monroe}}, \ and\ \bibinfo {author}
  {\bibfnamefont {D.~J.}\ \bibnamefont {Wineland}},\ }\href@noop {} {\bibfield
  {journal} {\bibinfo  {journal} {Nature}\ }\textbf {\bibinfo {volume} {409}},\
  \bibinfo {pages} {791} (\bibinfo {year} {2001})}\BibitemShut {NoStop}%
\bibitem [{\citenamefont {Ansmann}\ \emph {et~al.}(2009)\citenamefont
  {Ansmann}, \citenamefont {Wang}, \citenamefont {Bialczak}, \citenamefont
  {Hofheinz}, \citenamefont {Lucero}, \citenamefont {Neeley}, \citenamefont
  {O'Connell}, \citenamefont {Sank}, \citenamefont {Weides}, \citenamefont
  {Wenner} \emph {et~al.}}]{ansmann2009violation}%
  \BibitemOpen
  \bibfield  {author} {\bibinfo {author} {\bibfnamefont {M.}~\bibnamefont
  {Ansmann}}, \bibinfo {author} {\bibfnamefont {H.}~\bibnamefont {Wang}},
  \bibinfo {author} {\bibfnamefont {R.~C.}\ \bibnamefont {Bialczak}}, \bibinfo
  {author} {\bibfnamefont {M.}~\bibnamefont {Hofheinz}}, \bibinfo {author}
  {\bibfnamefont {E.}~\bibnamefont {Lucero}}, \bibinfo {author} {\bibfnamefont
  {M.}~\bibnamefont {Neeley}}, \bibinfo {author} {\bibfnamefont
  {A.}~\bibnamefont {O'Connell}}, \bibinfo {author} {\bibfnamefont
  {D.}~\bibnamefont {Sank}}, \bibinfo {author} {\bibfnamefont {M.}~\bibnamefont
  {Weides}}, \bibinfo {author} {\bibfnamefont {J.}~\bibnamefont {Wenner}},
  \emph {et~al.},\ }\href@noop {} {\bibfield  {journal} {\bibinfo  {journal}
  {Nature}\ }\textbf {\bibinfo {volume} {461}},\ \bibinfo {pages} {504}
  (\bibinfo {year} {2009})}\BibitemShut {NoStop}%
\bibitem [{\citenamefont {Rosenfeld}\ \emph {et~al.}(2017)\citenamefont
  {Rosenfeld}, \citenamefont {Burchardt}, \citenamefont {Garthoff},
  \citenamefont {Redeker}, \citenamefont {Ortegel}, \citenamefont {Rau},\ and\
  \citenamefont {Weinfurter}}]{rosenfeld2017event}%
  \BibitemOpen
  \bibfield  {author} {\bibinfo {author} {\bibfnamefont {W.}~\bibnamefont
  {Rosenfeld}}, \bibinfo {author} {\bibfnamefont {D.}~\bibnamefont
  {Burchardt}}, \bibinfo {author} {\bibfnamefont {R.}~\bibnamefont {Garthoff}},
  \bibinfo {author} {\bibfnamefont {K.}~\bibnamefont {Redeker}}, \bibinfo
  {author} {\bibfnamefont {N.}~\bibnamefont {Ortegel}}, \bibinfo {author}
  {\bibfnamefont {M.}~\bibnamefont {Rau}}, \ and\ \bibinfo {author}
  {\bibfnamefont {H.}~\bibnamefont {Weinfurter}},\ }\href@noop {} {\bibfield
  {journal} {\bibinfo  {journal} {Physical review letters}\ }\textbf {\bibinfo
  {volume} {119}},\ \bibinfo {pages} {010402} (\bibinfo {year}
  {2017})}\BibitemShut {NoStop}%
\bibitem [{\citenamefont {B{\"u}cker}\ \emph {et~al.}(2011)\citenamefont
  {B{\"u}cker}, \citenamefont {Grond}, \citenamefont {Manz}, \citenamefont
  {Berrada}, \citenamefont {Betz}, \citenamefont {Koller}, \citenamefont
  {Hohenester}, \citenamefont {Schumm}, \citenamefont {Perrin},\ and\
  \citenamefont {Schmiedmayer}}]{bucker2011twin}%
  \BibitemOpen
  \bibfield  {author} {\bibinfo {author} {\bibfnamefont {R.}~\bibnamefont
  {B{\"u}cker}}, \bibinfo {author} {\bibfnamefont {J.}~\bibnamefont {Grond}},
  \bibinfo {author} {\bibfnamefont {S.}~\bibnamefont {Manz}}, \bibinfo {author}
  {\bibfnamefont {T.}~\bibnamefont {Berrada}}, \bibinfo {author} {\bibfnamefont
  {T.}~\bibnamefont {Betz}}, \bibinfo {author} {\bibfnamefont {C.}~\bibnamefont
  {Koller}}, \bibinfo {author} {\bibfnamefont {U.}~\bibnamefont {Hohenester}},
  \bibinfo {author} {\bibfnamefont {T.}~\bibnamefont {Schumm}}, \bibinfo
  {author} {\bibfnamefont {A.}~\bibnamefont {Perrin}}, \ and\ \bibinfo {author}
  {\bibfnamefont {J.}~\bibnamefont {Schmiedmayer}},\ }\href@noop {} {\bibfield
  {journal} {\bibinfo  {journal} {Nature Physics}\ }\textbf {\bibinfo {volume}
  {7}},\ \bibinfo {pages} {608} (\bibinfo {year} {2011})}\BibitemShut {NoStop}%
\bibitem [{\citenamefont {Bonneau}\ \emph {et~al.}(2013)\citenamefont
  {Bonneau}, \citenamefont {Ruaudel}, \citenamefont {Lopes}, \citenamefont
  {Jaskula}, \citenamefont {Aspect}, \citenamefont {Boiron},\ and\
  \citenamefont {Westbrook}}]{BonneauPRA2013}%
  \BibitemOpen
  \bibfield  {author} {\bibinfo {author} {\bibfnamefont {M.}~\bibnamefont
  {Bonneau}}, \bibinfo {author} {\bibfnamefont {J.}~\bibnamefont {Ruaudel}},
  \bibinfo {author} {\bibfnamefont {R.}~\bibnamefont {Lopes}}, \bibinfo
  {author} {\bibfnamefont {J.-C.}\ \bibnamefont {Jaskula}}, \bibinfo {author}
  {\bibfnamefont {A.}~\bibnamefont {Aspect}}, \bibinfo {author} {\bibfnamefont
  {D.}~\bibnamefont {Boiron}}, \ and\ \bibinfo {author} {\bibfnamefont {C.~I.}\
  \bibnamefont {Westbrook}},\ }\href {\doibase 10.1103/PhysRevA.87.061603}
  {\bibfield  {journal} {\bibinfo  {journal} {Phys. Rev. A}\ }\textbf {\bibinfo
  {volume} {87}},\ \bibinfo {pages} {061603} (\bibinfo {year}
  {2013})}\BibitemShut {NoStop}%
\bibitem [{\citenamefont {Lopes}\ \emph {et~al.}(2015)\citenamefont {Lopes},
  \citenamefont {Imanaliev}, \citenamefont {Aspect}, \citenamefont {Cheneau},
  \citenamefont {Boiron},\ and\ \citenamefont {Westbrook}}]{lopes2015atomic}%
  \BibitemOpen
  \bibfield  {author} {\bibinfo {author} {\bibfnamefont {R.}~\bibnamefont
  {Lopes}}, \bibinfo {author} {\bibfnamefont {A.}~\bibnamefont {Imanaliev}},
  \bibinfo {author} {\bibfnamefont {A.}~\bibnamefont {Aspect}}, \bibinfo
  {author} {\bibfnamefont {M.}~\bibnamefont {Cheneau}}, \bibinfo {author}
  {\bibfnamefont {D.}~\bibnamefont {Boiron}}, \ and\ \bibinfo {author}
  {\bibfnamefont {C.~I.}\ \bibnamefont {Westbrook}},\ }\href@noop {} {\bibfield
   {journal} {\bibinfo  {journal} {Nature}\ }\textbf {\bibinfo {volume}
  {520}},\ \bibinfo {pages} {66} (\bibinfo {year} {2015})}\BibitemShut
  {NoStop}%
\bibitem [{\citenamefont {Dussarrat}\ \emph {et~al.}(2017)\citenamefont
  {Dussarrat}, \citenamefont {Perrier}, \citenamefont {Imanaliev},
  \citenamefont {Lopes}, \citenamefont {Aspect}, \citenamefont {Cheneau},
  \citenamefont {Boiron},\ and\ \citenamefont {Westbrook}}]{dussarrat2017two}%
  \BibitemOpen
  \bibfield  {author} {\bibinfo {author} {\bibfnamefont {P.}~\bibnamefont
  {Dussarrat}}, \bibinfo {author} {\bibfnamefont {M.}~\bibnamefont {Perrier}},
  \bibinfo {author} {\bibfnamefont {A.}~\bibnamefont {Imanaliev}}, \bibinfo
  {author} {\bibfnamefont {R.}~\bibnamefont {Lopes}}, \bibinfo {author}
  {\bibfnamefont {A.}~\bibnamefont {Aspect}}, \bibinfo {author} {\bibfnamefont
  {M.}~\bibnamefont {Cheneau}}, \bibinfo {author} {\bibfnamefont
  {D.}~\bibnamefont {Boiron}}, \ and\ \bibinfo {author} {\bibfnamefont {C.~I.}\
  \bibnamefont {Westbrook}},\ }\href@noop {} {\bibfield  {journal} {\bibinfo
  {journal} {Physical review letters}\ }\textbf {\bibinfo {volume} {119}},\
  \bibinfo {pages} {173202} (\bibinfo {year} {2017})}\BibitemShut {NoStop}%
\bibitem [{\citenamefont {Jaskula}\ \emph {et~al.}(2010)\citenamefont
  {Jaskula}, \citenamefont {Bonneau}, \citenamefont {Partridge}, \citenamefont
  {Krachmalnicoff}, \citenamefont {Deuar}, \citenamefont {Kheruntsyan},
  \citenamefont {Aspect}, \citenamefont {Boiron},\ and\ \citenamefont
  {Westbrook}}]{PhysRevLett.105.190402}%
  \BibitemOpen
  \bibfield  {author} {\bibinfo {author} {\bibfnamefont {J.-C.}\ \bibnamefont
  {Jaskula}}, \bibinfo {author} {\bibfnamefont {M.}~\bibnamefont {Bonneau}},
  \bibinfo {author} {\bibfnamefont {G.~B.}\ \bibnamefont {Partridge}}, \bibinfo
  {author} {\bibfnamefont {V.}~\bibnamefont {Krachmalnicoff}}, \bibinfo
  {author} {\bibfnamefont {P.}~\bibnamefont {Deuar}}, \bibinfo {author}
  {\bibfnamefont {K.~V.}\ \bibnamefont {Kheruntsyan}}, \bibinfo {author}
  {\bibfnamefont {A.}~\bibnamefont {Aspect}}, \bibinfo {author} {\bibfnamefont
  {D.}~\bibnamefont {Boiron}}, \ and\ \bibinfo {author} {\bibfnamefont {C.~I.}\
  \bibnamefont {Westbrook}},\ }\href {\doibase 10.1103/PhysRevLett.105.190402}
  {\bibfield  {journal} {\bibinfo  {journal} {Phys. Rev. Lett.}\ }\textbf
  {\bibinfo {volume} {105}},\ \bibinfo {pages} {190402} (\bibinfo {year}
  {2010})}\BibitemShut {NoStop}%
\bibitem [{\citenamefont {Shin}\ \emph {et~al.}(2019)\citenamefont {Shin},
  \citenamefont {Henson}, \citenamefont {Hodgman}, \citenamefont {Wasak},
  \citenamefont {Chwede{\'n}czuk},\ and\ \citenamefont
  {Truscott}}]{shin2019bell}%
  \BibitemOpen
  \bibfield  {author} {\bibinfo {author} {\bibfnamefont {D.}~\bibnamefont
  {Shin}}, \bibinfo {author} {\bibfnamefont {B.}~\bibnamefont {Henson}},
  \bibinfo {author} {\bibfnamefont {S.}~\bibnamefont {Hodgman}}, \bibinfo
  {author} {\bibfnamefont {T.}~\bibnamefont {Wasak}}, \bibinfo {author}
  {\bibfnamefont {J.}~\bibnamefont {Chwede{\'n}czuk}}, \ and\ \bibinfo {author}
  {\bibfnamefont {A.}~\bibnamefont {Truscott}},\ }\href@noop {} {\bibfield
  {journal} {\bibinfo  {journal} {Nature communications}\ }\textbf {\bibinfo
  {volume} {10}},\ \bibinfo {pages} {1} (\bibinfo {year} {2019})}\BibitemShut
  {NoStop}%
\bibitem [{\citenamefont {Kofler}\ \emph {et~al.}(2012)\citenamefont {Kofler},
  \citenamefont {Singh}, \citenamefont {Ebner}, \citenamefont {Keller},
  \citenamefont {Kotyrba},\ and\ \citenamefont
  {Zeilinger}}]{kofler2012einstein}%
  \BibitemOpen
  \bibfield  {author} {\bibinfo {author} {\bibfnamefont {J.}~\bibnamefont
  {Kofler}}, \bibinfo {author} {\bibfnamefont {M.}~\bibnamefont {Singh}},
  \bibinfo {author} {\bibfnamefont {M.}~\bibnamefont {Ebner}}, \bibinfo
  {author} {\bibfnamefont {M.}~\bibnamefont {Keller}}, \bibinfo {author}
  {\bibfnamefont {M.}~\bibnamefont {Kotyrba}}, \ and\ \bibinfo {author}
  {\bibfnamefont {A.}~\bibnamefont {Zeilinger}},\ }\href@noop {} {\bibfield
  {journal} {\bibinfo  {journal} {Physical Review A}\ }\textbf {\bibinfo
  {volume} {86}},\ \bibinfo {pages} {032115} (\bibinfo {year}
  {2012})}\BibitemShut {NoStop}%
\bibitem [{\citenamefont {Lewis-Swan}\ and\ \citenamefont
  {Kheruntsyan}(2015)}]{lewis2015proposal}%
  \BibitemOpen
  \bibfield  {author} {\bibinfo {author} {\bibfnamefont {R.~J.}\ \bibnamefont
  {Lewis-Swan}}\ and\ \bibinfo {author} {\bibfnamefont {K.}~\bibnamefont
  {Kheruntsyan}},\ }\href@noop {} {\bibfield  {journal} {\bibinfo  {journal}
  {Physical Review A}\ }\textbf {\bibinfo {volume} {91}},\ \bibinfo {pages}
  {052114} (\bibinfo {year} {2015})}\BibitemShut {NoStop}%
\bibitem [{\citenamefont {Bonneau}\ \emph {et~al.}(2018)\citenamefont
  {Bonneau}, \citenamefont {Munro}, \citenamefont {Nemoto},\ and\ \citenamefont
  {Schmiedmayer}}]{bonneau2018characterizing}%
  \BibitemOpen
  \bibfield  {author} {\bibinfo {author} {\bibfnamefont {M.}~\bibnamefont
  {Bonneau}}, \bibinfo {author} {\bibfnamefont {W.~J.}\ \bibnamefont {Munro}},
  \bibinfo {author} {\bibfnamefont {K.}~\bibnamefont {Nemoto}}, \ and\ \bibinfo
  {author} {\bibfnamefont {J.}~\bibnamefont {Schmiedmayer}},\ }\href@noop {}
  {\bibfield  {journal} {\bibinfo  {journal} {Physical Review A}\ }\textbf
  {\bibinfo {volume} {98}},\ \bibinfo {pages} {033608} (\bibinfo {year}
  {2018})}\BibitemShut {NoStop}%
\bibitem [{\citenamefont {Petrov}\ \emph {et~al.}(2000)\citenamefont {Petrov},
  \citenamefont {Shlyapnikov},\ and\ \citenamefont
  {Walraven}}]{petrov2000regimes}%
  \BibitemOpen
  \bibfield  {author} {\bibinfo {author} {\bibfnamefont {D.}~\bibnamefont
  {Petrov}}, \bibinfo {author} {\bibfnamefont {G.}~\bibnamefont {Shlyapnikov}},
  \ and\ \bibinfo {author} {\bibfnamefont {J.}~\bibnamefont {Walraven}},\
  }\href@noop {} {\bibfield  {journal} {\bibinfo  {journal} {Physical Review
  Letters}\ }\textbf {\bibinfo {volume} {85}},\ \bibinfo {pages} {3745}
  (\bibinfo {year} {2000})}\BibitemShut {NoStop}%
\bibitem [{\citenamefont {Folman}\ \emph {et~al.}(2000)\citenamefont {Folman},
  \citenamefont {Kr\"uger}, \citenamefont {Cassettari}, \citenamefont {Hessmo},
  \citenamefont {Maier},\ and\ \citenamefont {Schmiedmayer}}]{FolmanPRL2000}%
  \BibitemOpen
  \bibfield  {author} {\bibinfo {author} {\bibfnamefont {R.}~\bibnamefont
  {Folman}}, \bibinfo {author} {\bibfnamefont {P.}~\bibnamefont {Kr\"uger}},
  \bibinfo {author} {\bibfnamefont {D.}~\bibnamefont {Cassettari}}, \bibinfo
  {author} {\bibfnamefont {B.}~\bibnamefont {Hessmo}}, \bibinfo {author}
  {\bibfnamefont {T.}~\bibnamefont {Maier}}, \ and\ \bibinfo {author}
  {\bibfnamefont {J.}~\bibnamefont {Schmiedmayer}},\ }\href {\doibase
  10.1103/PhysRevLett.84.4749} {\bibfield  {journal} {\bibinfo  {journal}
  {Phys. Rev. Lett.}\ }\textbf {\bibinfo {volume} {84}},\ \bibinfo {pages}
  {4749} (\bibinfo {year} {2000})}\BibitemShut {NoStop}%
\bibitem [{\citenamefont {Schumm}\ \emph {et~al.}(2005)\citenamefont {Schumm},
  \citenamefont {Hofferberth}, \citenamefont {Andersson}, \citenamefont
  {Wildermuth}, \citenamefont {Groth}, \citenamefont {Bar-Joseph},
  \citenamefont {Schmiedmayer},\ and\ \citenamefont
  {Kr{\"{u}}ger}}]{Schumm2005b}%
  \BibitemOpen
  \bibfield  {author} {\bibinfo {author} {\bibfnamefont {T.}~\bibnamefont
  {Schumm}}, \bibinfo {author} {\bibfnamefont {S.}~\bibnamefont {Hofferberth}},
  \bibinfo {author} {\bibfnamefont {L.~M.}\ \bibnamefont {Andersson}}, \bibinfo
  {author} {\bibfnamefont {S.}~\bibnamefont {Wildermuth}}, \bibinfo {author}
  {\bibfnamefont {S.}~\bibnamefont {Groth}}, \bibinfo {author} {\bibfnamefont
  {I.}~\bibnamefont {Bar-Joseph}}, \bibinfo {author} {\bibfnamefont
  {J.}~\bibnamefont {Schmiedmayer}}, \ and\ \bibinfo {author} {\bibfnamefont
  {P.}~\bibnamefont {Kr{\"{u}}ger}},\ }\href {\doibase 10.1038/nphys125}
  {\bibfield  {journal} {\bibinfo  {journal} {Nature Physics}\ }\textbf
  {\bibinfo {volume} {1}},\ \bibinfo {pages} {57} (\bibinfo {year}
  {2005})}\BibitemShut {NoStop}%
\bibitem [{\citenamefont {Hofferberth}\ \emph {et~al.}(2006)\citenamefont
  {Hofferberth}, \citenamefont {Lesanovsky}, \citenamefont {Fischer},
  \citenamefont {Verdu},\ and\ \citenamefont
  {Schmiedmayer}}]{hofferberth2006radiofrequency}%
  \BibitemOpen
  \bibfield  {author} {\bibinfo {author} {\bibfnamefont {S.}~\bibnamefont
  {Hofferberth}}, \bibinfo {author} {\bibfnamefont {I.}~\bibnamefont
  {Lesanovsky}}, \bibinfo {author} {\bibfnamefont {B.}~\bibnamefont {Fischer}},
  \bibinfo {author} {\bibfnamefont {J.}~\bibnamefont {Verdu}}, \ and\ \bibinfo
  {author} {\bibfnamefont {J.}~\bibnamefont {Schmiedmayer}},\ }\href@noop {}
  {\bibfield  {journal} {\bibinfo  {journal} {Nature Physics}\ }\textbf
  {\bibinfo {volume} {2}},\ \bibinfo {pages} {710} (\bibinfo {year}
  {2006})}\BibitemShut {NoStop}%
\bibitem [{\citenamefont {Lesanovsky}\ \emph {et~al.}(2006)\citenamefont
  {Lesanovsky}, \citenamefont {Schumm}, \citenamefont {Hofferberth},
  \citenamefont {Andersson}, \citenamefont {Kr{\"u}ger},\ and\ \citenamefont
  {Schmiedmayer}}]{LesanovskyPRA2006}%
  \BibitemOpen
  \bibfield  {author} {\bibinfo {author} {\bibfnamefont {I.}~\bibnamefont
  {Lesanovsky}}, \bibinfo {author} {\bibfnamefont {T.}~\bibnamefont {Schumm}},
  \bibinfo {author} {\bibfnamefont {S.}~\bibnamefont {Hofferberth}}, \bibinfo
  {author} {\bibfnamefont {L.~M.}\ \bibnamefont {Andersson}}, \bibinfo {author}
  {\bibfnamefont {P.}~\bibnamefont {Kr{\"u}ger}}, \ and\ \bibinfo {author}
  {\bibfnamefont {J.}~\bibnamefont {Schmiedmayer}},\ }\href@noop {} {\bibfield
  {journal} {\bibinfo  {journal} {Physical Review A}\ }\textbf {\bibinfo
  {volume} {73}},\ \bibinfo {pages} {033619} (\bibinfo {year}
  {2006})}\BibitemShut {NoStop}%
  \bibitem [{1()}]{SM}%
  \BibitemOpen
  \href@noop {} {}\bibinfo {note} {See Supplemental Material at \url{here place link to SM} for
  more details on state inversion using optimal-control techniques, theoretical
  calculation of the expected two-particle state, possible extension to a fermionic system, the imaging setup and
  detection noise, which includes Refs.~\cite
  {basden2003photon,FrankNC2014,FrankSR2016,CanevaPRA2011,zhao2007high,bucker2013twin,pigneurnon}. }\BibitemShut
  {Stop}%
\bibitem [{\citenamefont {Gerbier}(2004)}]{gerbier2004quasi}%
  \BibitemOpen
  \bibfield  {author} {\bibinfo {author} {\bibfnamefont {F.}~\bibnamefont
  {Gerbier}},\ }\href@noop {} {\bibfield  {journal} {\bibinfo  {journal} {EPL
  (Europhysics Letters)}\ }\textbf {\bibinfo {volume} {66}},\ \bibinfo {pages}
  {771} (\bibinfo {year} {2004})}\BibitemShut {NoStop}%
\bibitem [{\citenamefont {Lewis-Swan}\ and\ \citenamefont
  {Kheruntsyan}(2020)}]{lewis2020atomic}%
  \BibitemOpen
  \bibfield  {author} {\bibinfo {author} {\bibfnamefont {R.}~\bibnamefont
  {Lewis-Swan}}\ and\ \bibinfo {author} {\bibfnamefont {K.}~\bibnamefont
  {Kheruntsyan}},\ }\href@noop {} {\bibfield  {journal} {\bibinfo  {journal}
  {Physical Review A}\ }\textbf {\bibinfo {volume} {101}},\ \bibinfo {pages}
  {043615} (\bibinfo {year} {2020})}\BibitemShut {NoStop}%
\bibitem [{\citenamefont {B{\"u}cker}\ \emph {et~al.}(2009)\citenamefont
  {B{\"u}cker}, \citenamefont {Perrin}, \citenamefont {Manz}, \citenamefont
  {Betz}, \citenamefont {Koller}, \citenamefont {Plisson}, \citenamefont
  {Rottmann}, \citenamefont {Schumm},\ and\ \citenamefont
  {Schmiedmayer}}]{bucker2009single}%
  \BibitemOpen
  \bibfield  {author} {\bibinfo {author} {\bibfnamefont {R.}~\bibnamefont
  {B{\"u}cker}}, \bibinfo {author} {\bibfnamefont {A.}~\bibnamefont {Perrin}},
  \bibinfo {author} {\bibfnamefont {S.}~\bibnamefont {Manz}}, \bibinfo {author}
  {\bibfnamefont {T.}~\bibnamefont {Betz}}, \bibinfo {author} {\bibfnamefont
  {C.}~\bibnamefont {Koller}}, \bibinfo {author} {\bibfnamefont
  {T.}~\bibnamefont {Plisson}}, \bibinfo {author} {\bibfnamefont
  {J.}~\bibnamefont {Rottmann}}, \bibinfo {author} {\bibfnamefont
  {T.}~\bibnamefont {Schumm}}, \ and\ \bibinfo {author} {\bibfnamefont
  {J.}~\bibnamefont {Schmiedmayer}},\ }\href@noop {} {\bibfield  {journal}
  {\bibinfo  {journal} {New Journal of Physics}\ }\textbf {\bibinfo {volume}
  {11}},\ \bibinfo {pages} {103039} (\bibinfo {year} {2009})}\BibitemShut
  {NoStop}%
\bibitem [{\citenamefont {Pigneur}\ \emph {et~al.}(2018)\citenamefont
  {Pigneur}, \citenamefont {Berrada}, \citenamefont {Bonneau}, \citenamefont
  {Schumm}, \citenamefont {Demler},\ and\ \citenamefont
  {Schmiedmayer}}]{pigneur2018relaxation}%
  \BibitemOpen
  \bibfield  {author} {\bibinfo {author} {\bibfnamefont {M.}~\bibnamefont
  {Pigneur}}, \bibinfo {author} {\bibfnamefont {T.}~\bibnamefont {Berrada}},
  \bibinfo {author} {\bibfnamefont {M.}~\bibnamefont {Bonneau}}, \bibinfo
  {author} {\bibfnamefont {T.}~\bibnamefont {Schumm}}, \bibinfo {author}
  {\bibfnamefont {E.}~\bibnamefont {Demler}}, \ and\ \bibinfo {author}
  {\bibfnamefont {J.}~\bibnamefont {Schmiedmayer}},\ }\href@noop {} {\bibfield
  {journal} {\bibinfo  {journal} {Physical review letters}\ }\textbf {\bibinfo
  {volume} {120}},\ \bibinfo {pages} {173601} (\bibinfo {year}
  {2018})}\BibitemShut {NoStop}%
\bibitem [{\citenamefont {Basden}\ \emph {et~al.}(2003)\citenamefont {Basden},
  \citenamefont {Haniff},\ and\ \citenamefont {Mackay}}]{basden2003photon}%
  \BibitemOpen
  \bibfield  {author} {\bibinfo {author} {\bibfnamefont {A.}~\bibnamefont
  {Basden}}, \bibinfo {author} {\bibfnamefont {C.}~\bibnamefont {Haniff}}, \
  and\ \bibinfo {author} {\bibfnamefont {C.}~\bibnamefont {Mackay}},\
  }\href@noop {} {\bibfield  {journal} {\bibinfo  {journal} {Monthly notices of
  the royal astronomical society}\ }\textbf {\bibinfo {volume} {345}},\
  \bibinfo {pages} {985} (\bibinfo {year} {2003})}\BibitemShut {NoStop}%
\bibitem [{\citenamefont {van Frank}\ \emph {et~al.}(2014)\citenamefont {van
  Frank}, \citenamefont {Negretti}, \citenamefont {Berrada}, \citenamefont
  {B{\"u}cker}, \citenamefont {Montangero}, \citenamefont {Schaff},
  \citenamefont {Schumm}, \citenamefont {Calarco},\ and\ \citenamefont
  {Schmiedmayer}}]{FrankNC2014}%
  \BibitemOpen
  \bibfield  {author} {\bibinfo {author} {\bibfnamefont {S.}~\bibnamefont {van
  Frank}}, \bibinfo {author} {\bibfnamefont {A.}~\bibnamefont {Negretti}},
  \bibinfo {author} {\bibfnamefont {T.}~\bibnamefont {Berrada}}, \bibinfo
  {author} {\bibfnamefont {R.}~\bibnamefont {B{\"u}cker}}, \bibinfo {author}
  {\bibfnamefont {S.}~\bibnamefont {Montangero}}, \bibinfo {author}
  {\bibfnamefont {J.-F.}\ \bibnamefont {Schaff}}, \bibinfo {author}
  {\bibfnamefont {T.}~\bibnamefont {Schumm}}, \bibinfo {author} {\bibfnamefont
  {T.}~\bibnamefont {Calarco}}, \ and\ \bibinfo {author} {\bibfnamefont
  {J.}~\bibnamefont {Schmiedmayer}},\ }\href@noop {} {\bibfield  {journal}
  {\bibinfo  {journal} {Nature communications}\ }\textbf {\bibinfo {volume}
  {5}},\ \bibinfo {pages} {1} (\bibinfo {year} {2014})}\BibitemShut {NoStop}%
\bibitem [{\citenamefont {van Frank}\ \emph {et~al.}(2016)\citenamefont {van
  Frank}, \citenamefont {Bonneau}, \citenamefont {Schmiedmayer}, \citenamefont
  {Hild}, \citenamefont {Gross}, \citenamefont {Cheneau}, \citenamefont
  {Bloch}, \citenamefont {Pichler}, \citenamefont {Negretti}, \citenamefont
  {Calarco} \emph {et~al.}}]{FrankSR2016}%
  \BibitemOpen
  \bibfield  {author} {\bibinfo {author} {\bibfnamefont {S.}~\bibnamefont {van
  Frank}}, \bibinfo {author} {\bibfnamefont {M.}~\bibnamefont {Bonneau}},
  \bibinfo {author} {\bibfnamefont {J.}~\bibnamefont {Schmiedmayer}}, \bibinfo
  {author} {\bibfnamefont {S.}~\bibnamefont {Hild}}, \bibinfo {author}
  {\bibfnamefont {C.}~\bibnamefont {Gross}}, \bibinfo {author} {\bibfnamefont
  {M.}~\bibnamefont {Cheneau}}, \bibinfo {author} {\bibfnamefont
  {I.}~\bibnamefont {Bloch}}, \bibinfo {author} {\bibfnamefont
  {T.}~\bibnamefont {Pichler}}, \bibinfo {author} {\bibfnamefont
  {A.}~\bibnamefont {Negretti}}, \bibinfo {author} {\bibfnamefont
  {T.}~\bibnamefont {Calarco}},  \emph {et~al.},\ }\href@noop {} {\bibfield
  {journal} {\bibinfo  {journal} {Scientific reports}\ }\textbf {\bibinfo
  {volume} {6}},\ \bibinfo {pages} {34187} (\bibinfo {year}
  {2016})}\BibitemShut {NoStop}%
\bibitem [{\citenamefont {Caneva}\ \emph {et~al.}(2011)\citenamefont {Caneva},
  \citenamefont {Calarco},\ and\ \citenamefont {Montangero}}]{CanevaPRA2011}%
  \BibitemOpen
  \bibfield  {author} {\bibinfo {author} {\bibfnamefont {T.}~\bibnamefont
  {Caneva}}, \bibinfo {author} {\bibfnamefont {T.}~\bibnamefont {Calarco}}, \
  and\ \bibinfo {author} {\bibfnamefont {S.}~\bibnamefont {Montangero}},\
  }\href {\doibase 10.1103/PhysRevA.84.022326} {\bibfield  {journal} {\bibinfo
  {journal} {Phys. Rev. A}\ }\textbf {\bibinfo {volume} {84}},\ \bibinfo
  {pages} {022326} (\bibinfo {year} {2011})}\BibitemShut {NoStop}%
\bibitem [{\citenamefont {B{\"u}cker}(2013)}]{bucker2013twin}%
  \BibitemOpen
  \bibfield  {author} {\bibinfo {author} {\bibfnamefont {R.}~\bibnamefont
  {B{\"u}cker}},\ }\emph {\bibinfo {title} {Twin-atom beam generation in a
  one-dimensional Bose gas}},\ \href@noop {} {Ph.D. thesis},\ \bibinfo
  {school} {Techische Universit\"at Wien} (\bibinfo {year} {2013})\BibitemShut
  {NoStop}%
\bibitem [{\citenamefont {Pigneur}(2020)}]{pigneurnon}%
  \BibitemOpen
  \bibfield  {author} {\bibinfo {author} {\bibfnamefont {M.}~\bibnamefont
  {Pigneur}},\ }\href@noop {} {\emph {\bibinfo {title} {Non-equilibrium
  Dynamics of Tunnel-Coupled Superfluids: Relaxation to a Phase-Locked
  Equilibrium State in a One-Dimensional Bosonic Josephson Junction}}}\
  (\bibinfo  {publisher} {Springer Nature},\ \bibinfo {year}
  {2020})\BibitemShut {NoStop}%
\bibitem [{\citenamefont {Zhao}\ \emph {et~al.}(2007)\citenamefont {Zhao},
  \citenamefont {Chen}, \citenamefont {Pan}, \citenamefont {Schmiedmayer},
  \citenamefont {Recati}, \citenamefont {Astrakharchik},\ and\ \citenamefont
  {Calarco}}]{zhao2007high}%
  \BibitemOpen
  \bibfield  {author} {\bibinfo {author} {\bibfnamefont {B.}~\bibnamefont
  {Zhao}}, \bibinfo {author} {\bibfnamefont {Z.-B.}\ \bibnamefont {Chen}},
  \bibinfo {author} {\bibfnamefont {J.-W.}\ \bibnamefont {Pan}}, \bibinfo
  {author} {\bibfnamefont {J.}~\bibnamefont {Schmiedmayer}}, \bibinfo {author}
  {\bibfnamefont {A.}~\bibnamefont {Recati}}, \bibinfo {author} {\bibfnamefont
  {G.~E.}\ \bibnamefont {Astrakharchik}}, \ and\ \bibinfo {author}
  {\bibfnamefont {T.}~\bibnamefont {Calarco}},\ }\href@noop {} {\bibfield
  {journal} {\bibinfo  {journal} {Physical Review A}\ }\textbf {\bibinfo
  {volume} {75}},\ \bibinfo {pages} {042312} (\bibinfo {year}
  {2007})}\BibitemShut {NoStop}%
\end{thebibliography}
%

\end{document}


\title{Supplemental Material: Two-particle Interference with Double Twin-atom Beams}
    
    \author{F. Borselli}
    \author{M. Maiwöger}
    \author{T. Zhang}
    \author{P. Haslinger}
    \affiliation{Vienna Center for Quantum Science and Technology, Atominstitut, TU Wien, 
    1020 Vienna, Austria}
    
    \author{V. Mukherjee}
    \affiliation{Indian Institute of Science Education and Research, 760010 Berhampur, India}
    
    \author{A. Negretti}
    \affiliation{The Hamburg Centre for Ultrafast Imaging, Universität Hamburg, D-22761 Hamburg, Germany}
    
    \author{S. Montangero}
    \affiliation{Dipartimento di Fisica e Astronomia “G. Galilei”, Università di Padova, I-35131 Padova, Italy} \affiliation{INFN Sezione di Padova, I-35131 Padua, Italy.}
    
    \author{T. Calarco}
    \affiliation{Forschungszentrum Jülich, Wilhelm-Johnen-Straße, D-52425 Jülich, 
    and University of Cologne, Institute for Theoretical Physics, D-50937 Cologne, Germany}
    
    \author{I. Mazets}
    \affiliation{Vienna Center for Quantum Science and Technology, Atominstitut, TU Wien, 
    1020 Vienna, Austria}
    \affiliation{Research Platform MMM ``Mathematics--Magnetism--Materials'', 
       c/o Fakult{\"a}t f{\"ur} Mathematik, Universit{\"a}t Wien, 1090 Vienna, Austria}

    \author{M. Bonneau}
    \author{J. Schmiedmayer}
    \affiliation{Vienna Center for Quantum Science and Technology, Atominstitut, TU Wien, 
    1020 Vienna, Austria}

\date{\today} 

\maketitle

\section{State inversion using optimal control techniques}
    The system consists of a quasi-one-dimensional condensate, i.e. a weakly interacting bosonic ensemble that is loosely confined longitudinally, but tightly confined transversally, as in previously realised optimal control experiments with atom chips~\cite{FrankNC2014,FrankSR2016}. In the transverse direction that hereafter we denote as the $y$-axis the potential is initially a single (anharmonic) well, as in Refs.~\cite{FrankNC2014,FrankSR2016}, but then it is controlled dynamically by means of an external radio-frequency field in order to transform it to a double-well potential~\cite{LesanovskyPRA2006}. As in previous related experiments~\cite{FrankNC2014,FrankSR2016}, the system dynamics along the $y$-axis can be described through an effective one-dimensional Gross-Pitaevskii equation, whose nonlinear Hamiltonian is given by 
            \begin{align}
                \label{eq:Hgp}
                \hat H_{\mathrm{gp}}[\psi, t] & = -\frac{\hbar^2}{2m} \frac{\partial^2}{\partial y^2} + V(y, t) + gN|\psi(y, t)|^2.
            \end{align}
    Here, $m$ is the mass of the boson, specifically of the alkali atom $^{87}$Rb, $V(y, t)$ is the time-dependent potential that we manipulate optimally, $g$ is the effective one-dimensional boson-boson coupling constant (see Ref.~\cite{FrankSR2016} for further details), $N$ is the number of bosons, and $\psi(y, t)$ is the condensate wavefunction formalised to unity. We note that because of the large separation of time scales between the transverse and longitudinal degrees of freedom, the quantum dynamics of the latter can be effectively assumed to be frozen during the excitation process in the transverse direction, which we are interested in. 

    The external potential $V(y, t)$ produced by the atom chip is approximated by
        \ba 
            V(y, t) &=&   a_0(t) + a_2(t)y^2 + a_4(t)y^4  + a_6(t)y^6, \non\\
            a_n(t) &=& \sum_{j=1}^6 \alpha_j^{(n)} [R_f(t)]^j~\text{for}~n = 0, 2, 4, 6,
        \ea
    where the time-independent parameters $\alpha_j^{(n)}$, which are provided in Tab.~\ref{tab:tabalpha}, have units of kHz/m$^{n}$. 
    The numerical values of the parameters $\alpha_j^{(n)}$ have been obtained by numerically fitting the simulated and experimentally calibrated potential generated by the atom chip with a polynomial of sixth order. This strategy has been adopted to simplify the numerical effort of the optimisation. The dimensionless time-dependent function $R_f(t)$ is proportional to the strength of the radio-frequency field applied to the atom chip and it is the control parameter we have to optimise.

  \begin{table}[hbt]
    \begin{tabular}{|m{2em}|m{5em}|m{5em}|m{5em}|m{5em}|} 
    \hline
    $~j~$ & $\alpha_j^{(0)}$ & $\alpha_j^{(2)}$ & $\alpha_j^{(4)}$ & $\alpha_j^{(6)}$ \\
    \hline
    ~0~& 54.451 & 74.025  & -3.4221 & 0.2406\\
    \hline
    ~1~& -8.6264 & -19.429 & 24.648 & -6.0581 \\ 
    \hline
    ~2~& 3570.3 & - 3309.1 & 1231.6 & -153.85\\ 
    \hline
    ~3~& -12650 & 18497 & -8450.8 & 1221.2\\ 
    \hline
    ~4~& 25646 & -46369 & 23425 & -3661.4\\ 
    \hline
    ~5~& -27546 & 56311 & -30416 & 5049.5\\ 
    \hline
    ~6~& 12106 & -26894 & 15268 &  -2663.3\\ 
    \hline
    \end{tabular}
    \caption{The parameters $\alpha_j^{(n)}$ in units of kHz/$\mu$m$^{n}$ for $n=0,2,4,6$.} 
    \label{tab:tabalpha}
    \end{table}
    
\begin{figure}[hbt]
        \centering
        \includegraphics[width=0.45\textwidth]{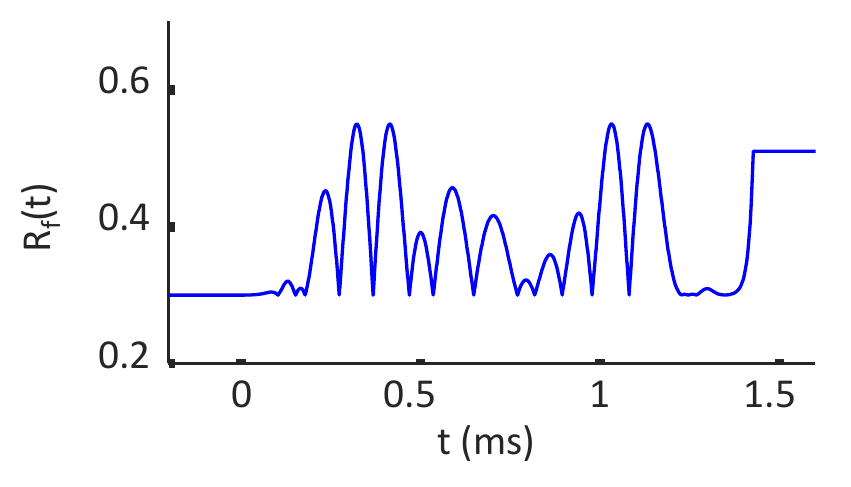}
        \caption{The ramp $R_{f}(t)$ of the amplitude of the radio-frequency field against time.}
        \label{fig:ramp}
    \end{figure}

    In the present experiment, the quasi-condensate is initially prepared in the ground state, $\varphi_0(y)$, of the initial single well potential $V(y,0)$. Our goal is to bring the quasi-condensate in the second excited state, $\varphi_2(y)$, of the external potential $V(y,t_f)$ in double-well configuration in a time $t_f$ shorter than the decoherence time of the system. Here, the nonlinear eigenstates $\varphi_{0,2}(y)$ of the Hamiltonian~(\ref{eq:Hgp}) are determined numerically by the imaginary-time technique with $N=700$. To this end, we employ optimal control techniques to generate the optimal radio-frequency field $R_f(t)$ that minimises the cost function defined at the final time $t_f$ as 
    \ba
    \mathcal{J} = 1 - \left\vert\int_{\mathbb{R}}\mathrm{d}y\,\varphi^*_2(y)\psi(y,t_f)\right\vert^2.
    \ea
   Specifically, we employ the CRAB optimisation method~\cite{CanevaPRA2011}. Here, the radio-frequency field $R_f(t)$ is expanded into a (not necessarily orthogonal) truncated basis
    \begin{widetext}
    \be
    \label{eq:rf}
    R_f(t) = 0.3 + \frac{1}{\lambda(t) N_f}\left|\sum_{j = 1}^{N_f}\left(c_j\cos\frac{2\pi f_j t}{t_f} + d_j \sin\frac{2\pi f_j t}{t_f} \right)\right| +  0.21\, e^{-8(t_f - t)},
    \ee
    \end{widetext}
    for $0 \leq t \leq t_f$. Here $N_f = 10$ denotes the total number of frequencies considered in Eq. \eqref{eq:rf}; the multiple frequencies allow us to engineer non-trivial pulses with multiple maxima and minima, as shown in Fig. \ref{fig:ramp}. The dimensionless function
    \begin{eqnarray}
        \lambda(t) = 0.5 + 10^4 \left[e^{-8t} + e^{-8(t_f - t)} \right]
    \end{eqnarray}
    is large and positive at $t=0,\,t_f$, thereby fixing the initial and final values of the RF-field. 
    On the other hand, $\lambda(t)$ assumes the value $0.5$ at intermediate times, so as to allow for variations of the RF-field within the interval $(0,t_f)$. Furthermore, owing to experimental constraints, we impose the condition $0.3 \leq R_f(t) \leq 0.55~\forall~t$. We note that the field~(\ref{eq:rf}) is already given in dimensionless units, where times are rescaled with respect to $1/\omega_0$. The optimisation is carried out by varying the parameters $c_j$, $d_j$ and $f_j$. Thus, the optimisation has been performed in such a way that the double-well potential $V(y, t_f)$ is obtained by setting $R_f( t_f) = 0.51$ at final time $t_f/\omega_0 = 1.4$ ms, while $R_f(0) = 0.3$ results in the initial single-well potential. The exponential function appearing in Eq.~\ref{eq:rf} and its width $1/8$ have been chosen such that it increases smoothly and monotonically to the numerical value 0.21 as $t \to t_f^{-}$, such that the control parameter reaches the target value $R_f(t_f) = 0.51$ and we avoid excitation of the condensate along the vertical z-axis. In Fig.~\ref{fig:ramp} the optimised curve of the parameter $R_f(t)$ is plotted against time. The values of $R_f(t)$ for $t < 0$ and $t > t_f = 1.4$ ms in Fig. \ref{fig:ramp} signify that $R_f(t)$ is time-dependent only for the intermediate optimization times $(0, t_f)$, while it assumes constant values outside this time-interval.

\subsection{Transfer efficiency}        

    We estimate the percentage of atoms transferred to the source state from the evolution of the wavefunction of the BEC after the excitation pulse. If more than one eigenstate of the potential are populated, we should observe a beating pattern in the momentum distribution varying with the holding time in the trap. If the excited wavefunction corresponds to the source state, which is an eigenstate of the double-well potential, the outcome would be a constant profile. 
    The experimental profile was fitted with a linear combination $\Psi_{guess}(y)$ of different single-particle eigenstates $\psi_{i}(y)$ up to the sixth order ($i=6$):
\begin{equation}    
    \Psi_{guess}(y)={p_2}\psi_{2}(y)+\sum_{i}e^{i\phi_i}\sqrt{p_i}\psi_i(y)
\end{equation}
    where $\phi_i$ ($i=0,1,4,6$) are the relative phases and $p_i$ ($i=0,1,2,4,6$) are the normalized contributions from the five different states considered. The odd components from the third and fifth order were excluded from the fit function based on symmetry arguments to reduce the number of free parameters. This is consistent with the transverse symmetry of the experimental data. 
    The main contribution to the experimental profile comes from the second excited state of the double-well potential (\(\sim\) 97\%), corresponding to the source state. This demonstrates the state inversion using the optimal control engineered sequence.

\section{Creation of the DTB state}
A wave function $\Psi (\mathbf{r}_1,\, \mathbf{r}_2)\equiv \Psi (x_1, \, y_1, \, z_1;\, x_2, \, y_2, \, z_2)$ of two 
spin-polarized bosonic atoms is symmetric with respect to the permutation of the coordinates $(x_1, \, y_1, \, z_1)$ and $(x_2, \, y_2, \, z_2)$ of these atoms. If we factorize it into the longitudinal $(\Vert )$ and transverse $(\perp )$ parts, $\Psi (x_1, \, y_1, \, z_1;\, x_2, \, y_2, \, z_2)=\Psi _\Vert (x_1;\, x_2)\Psi _\perp (y_1, \, z_1;\, y_2, \, z_2)$ we readily see that each of them must be symmetric with respect to the permutation of its coordinates, $\Psi _\Vert (x_1;\, x_2)=\Psi _\Vert (x_2;\, x_1)$ and $\Psi _\perp (y_1, \, z_1;\, y_2, \, z_2)=\Psi _\perp (y_2, \, z_2;\, y_1, \, z_1)$, in order to be non-zero at $\mathbf{r}_1= \mathbf{r}_2$, thus allowing for $s$-wave scattering on the two-atom contact interaction (pseudo)potential $\propto \delta (x_1-x_2)\delta (y_1-y_2)\delta (z_1-z_2)$, where $\delta $ is the Dirac delta-function. 
We can write the initial state of the two particles in the spatial basis as
\begin{widetext}
\be
\Psi_{in}(x_1, \, y_1, \, z_1;\, x_2, \, y_2, \, z_2)=\langle y_1, \, z_1 | 2 \rangle \langle y_2, \,  z_2 | 2 \rangle \langle x_1 | k_x = 0 \rangle \langle x_2 | k_x = 0 \rangle,
\ee
\end{widetext}
where $\ket{n_y}=\ket{2}$ represents the transverse second-excited state and we assumed that longitudinally only the $\ket{k_x=0}$ mode is initially populated (the source state is at rest longitudinally).

The emission process conserves the symmetry of the initial two-particle wavefunction. However, the $\delta (x_1-x_2)$ term precludes transitions into antisymmetric longitudinal states. Therefore, only the longitudinal state $(\langle x_1 | -k_0 \rangle \langle x_2 | k_0 \rangle + \langle x_2 | -k_0 \rangle \langle x_1 | k_0 \rangle)/\sqrt{2}$ is possible. 
For the transverse component of the final wavefunction, let us consider the $\{ |L\rangle, \, |R\rangle \}$ basis. Due to bosonic symmetry, the transverse state is also symmetric with respect to the exchange of transverse coordinates.
However, transitions to $|L R \rangle$ or $|R L \rangle$ are not possible, since 

\be
\begin{split}
\int dy_1\int dy_2 \int dz_1\int dz_2\, \langle 2|y_1, \, z_1 \rangle \langle 2|y_2, \, z_2 \rangle   \delta (y_1-y_2)\delta (z_1-z_2) & \frac {\langle y_1, \, z_1|L \rangle \langle  y_2, \, z_2|R \rangle +
\langle y_1, \, z_1|R \rangle \langle y_2, \, z_2|L \rangle }{\sqrt{2}} \\
= \sqrt{2} \int dy\int dz\, (\langle 2|y,\, z \rangle )^2 \langle y, \, z|L\rangle \langle y, \, z|R \rangle\, =0 
\end{split} 
\ee


due to negligible overlap between the states $|L\rangle $ and $|R\rangle $. 

The remaining symmetric transverse states are $\ket{\Psi_{\perp}^-}=(\ket{LL}-\ket{RR})/\sqrt{2}$ and $\ket{\Psi_{\perp}^+}=(\ket{LL}+\ket{RR})/\sqrt{2}$.
Since for a perfectly symmetric trap
\be
\int dy\int dz\, (\langle 2|y,\, z \rangle )^2 (\langle y, \, z|L\rangle )^2 \, = \int dy\int dz\, (\langle 2|y,\, z \rangle )^2 (\langle y, \, z|R\rangle )^2,
\ee    
the matrix element for a transition from $|2\rangle |2 \rangle $ to $\ket{\Psi_{\perp}^-}$
vanishes due to destructive interference. 
The only non-zero matrix element couples $|2\rangle |2\rangle $ to 
$\ket{\Psi_{\perp}^+} $. Taking into account also the longitudinal component  
$(e^{-ik(x_1-x_2)}+ e^{-ik(x_2-x_1)})/\sqrt{2}$, we recover the DTB state $ \ket{\Psi_{DTB}} = (|L\rangle _-|L\rangle _+ + |R\rangle _-|R\rangle _+)/\sqrt{2}$.

\subsection{Extension to a fermionic system }

In our present experiment the source state from where the atom pairs are emitted relies on a Bose-Einstein condensate which has a defined longitudinal momentum $k_x=0$.  Moreover, the emitted double twin-atom beams are created by an $s$-wave scattering process. The same procedure does not apply to a fermionic gas. 
The atoms in a fermionic source state would have many longitudinal momenta up to the Fermi momentum $k_F$ and the total momentum of the emitted atom pairs would be not well defined. Moreover, spin-polarized fermions do not experience $s$-wave scattering, hence the collisional process at the basis of the emission of twin-beams would be completely different. So a source of fermionic twin atoms would have to look completely different. One can imagine breaking up a bosonic diatomic Feschbach molecule into its fermionic components as for example proposed as source for entangled atom pairs in~\cite{zhao2007high}, but imprinting a significant momentum on them would require additional processes like transferring the molecule before the break-up into a higher excited quasi bound state.
We could then envision such a system that produces twin fermionic atoms in a single waveguide. The spin degree of freedom would replace the double waveguide transverse degree of freedom of our setup and the emitted state would be a maximally entangled spin state $|\Phi^- \rangle = (\ket{\downarrow}_-\ket{\uparrow}_+ - \ket{\uparrow}_-\ket{\downarrow}_+)/\sqrt{2}$.


\section{Twin character and total transverse squeezing}

As already done in \cite{bucker2011twin}, we check the twin character of the DTB emission by looking at the fluctuations of the difference photon number $S_-$ between the atoms with momentum $\pm k_0$ over the different experimental realizations. 
For the separation data, we simply integrate over the two transverse modes ${S}_-=({S}_{L_-}+{S}_{R_-}) - ({S}_{L_+}+{S}_{R_+})$, where ${S}_{L_-}$ is the signal contained in the black box $L_-$ in Fig.~2a corresponding to the single-particle mode $\ket{L_-}$ (and similarly for the others). If there is no correlation among the signals in the two zones that are being analysed, the signal difference follows a binomial distribution. 
We can then evaluate the number squeezing factor $\xi^2_{k_0/-k_0}$ between the two longitudinal momentum classes and classify $\xi^2_{k_0/-k_0} < 1$ as a number-squeezed emission.
The main information about the data are listed in Tab.~\ref{tab:table1}. In particular, the results on the noise-corrected $\xi^2_{k_0/-k_0}$ between the two momentum states $\pm k_0$ confirm the results in \cite{bucker2011twin}, thus demonstrating the presence of a strongly non-Poissonian amount of correlations between the DTBs of opposite momenta. The error on $\xi^2$ is estimated using a bootstrapping method comparing 50 statistical copies of the full experiment.
\begin{table}
        \centering
        \begin{tabular}{|c|c|c|c|}
            \hline
             &          
            \multicolumn{2}{l}{\bfseries Separation} & 
            \multicolumn{1}{|c|}{\bfseries Interference}\\ 
            \hline
            $t_{hold} (\SI{}{\ms})$ & 0.025 & 0.425 & 0.025 \\
            Total images & 684 & 825 & 1498 \\
            $\tilde{\xi}^2_{k_0/-k_0}$ & 0.18 & 0.22 & 0.30\\
            $\xi^2_{k_0/-k_0}$ & 0.10(1) & 0.14(1) & 0.10(1)\\
            \hline
        \end{tabular}
        \caption{\textbf{Twin character} Main parameters of the two sets of data considered in this paper: the uncorrected atom number-squeezing factor $\tilde{\xi}^2_{k_0/-k_0}$ and the noise-corrected one $\xi^2_{k_0/-k_0}$. The latter is an indicator of the twin character of the DTB emission, i.e. the process of creation of pairs of atoms carrying opposite momenta.}
        \label{tab:table1}
\end{table}        

In the separation procedure we can also consider the total transverse number squeezing, i.e. the signal difference between the number of pairs emitted in the L- and in the R-waveguide, after integrating on the two longitudinal momenta (see light-blue dashed boxes in Fig.~2b). 
Since the atoms are detected pairwise independently into the L- and R-waveguide, we expect the distribution of the pairs difference $M_-$ to be binomial (uncorrelated). Hence, if $M_+$ is the total number of pairs, the corresponding variance is $\Delta M_-^2 = M_+$. 
Let us now consider the distribution of the signal difference $N_-$ between the atoms in the L-waveguide and R-waveguide and its variance $\Delta N_-^2$.
Since $var(aX)=a^2 var(X)$ for any variable $X$ where $a$ is a constant, we get $\Delta N_-^2 = \Delta^2(2 M_- ) = 4 \Delta N_-^2=4M_+=2N_+$.
From this follows that $\xi^2_{L/R} \equiv \Delta N_-^2/\Delta_b N_-^2=(2N_+)/N_+ =2$.


\section{Imaging system}
 
    Our fluorescence based imaging system consists of a nearly resonant sheet of light made of two counter-propagating laser beams. The light-sheet excites the atoms and make them undergo several absorption-spontaneous emission cycles. Part of these photons are collected on a camera placed below the atom chip and converted into electrons. 
    In principle, single atom recognition is possible and was already demonstrated in this system \cite{bucker2009single}.
    If the two counter-propagating laser beams are not exactly overlapped, or if their power is unbalanced, a light-pressure effect can show up in the fluorescence picture. For the data considered in this paper this effect is only residual (Fig.~2a).
    We take pictures after a time-of-flight $t_{TOF} = \SI{44}{\ms}$.

    \subsection{Far-field regime for the transverse direction}
    The final position $x_F$ of a trapped particle along a certain spatial direction $x$ after its release from the trap at time $t=0$ and a time-of-flight $t_{TOF} \equiv t_F$ reads $x_F = x_0+  \dot{x}_0 \cdot t_F= x_0 + p_0/m \cdot t_F$, where $x_0$  ($\dot{x}_0 =  p_0/m$) represents the initial position (velocity) of the particle in the trap at the moment of the release and $p_0$ its initial momentum. Assuming a harmonic trapping ($\dot{x} =i\omega_x x$) with angular frequency $\omega_x$ along the x-axis, we derive the expression $x_F =x_0+i\omega_x x_0\cdot t_F=x_0+p_0/m\cdot t_F$. The condition of the final position expressing the initial momentum of the particle in the trap at the moment of the release is then $p_0/m \cdot t_F \gg x_0$, which translates into the requirement $i\omega_x x_0 \cdot t_F \gg x_0 \rightarrow t_F \gg 1/\omega_{x,y,z}$, independently of the spatial direction we are referring to.
    In our experimental setup we have $\omega_x \simeq 2\pi \cdot 10$ Hz and $\omega_{y,z} = 2\pi \cdot 2$ kHz, which corresponds to $1/\omega_x \simeq 16$ ms and $1/\omega_{y,z} \simeq 0.1$ ms. Since in our setup $t_F=44$ ms,
    the condition $t_F \gg 1/\omega_{x,y,z}$ is well satisfied along the transverse y,z-axis and only partly satisfied along the x-axis. This shows that the transverse expansion of the atomic cloud after its release from the chip trap is fast compared to $t_{TOF}$, hence the fluorescence image of the final cloud shows the in-situ momentum distribution along the y-axis.

\subsection{Atom detection and detection noise}
    Experimentally we cannot access the atom number directly, but rather we measure the number of photons hitting the camera.
    Having considered two boxes 1 and 2 on a typical fluorescence image (consider for example any two black boxes in Fig.~2a), we define the sum and difference photon signal relative to the two boxes $S_{\pm}=S_1-S_2$, where $S_1$ ($S_2$) is the measured fluorescence signal from box 1 (2).
    If we assume that to each imaged atom correspond exactly $p$ photons, then we can write
        \be
        S_{\pm} = p N_{\pm},
        \label{eq:ppa}
        \ee
    where $N_{\pm}$ is the sum or difference atom number relative to the same two boxes.
    Having assumed $p$ constant, we can derive an expression for the variance of the signal difference $\Delta S_-^2 \equiv \text{var}(pN_-)=p^2\Delta N_-^2$.
    Using Eq.~\ref{eq:ppa} and the expression $\Delta_{b} N_-^2=N_+$ we get
    \be
    \tilde{\xi}^2 \equiv \frac{\Delta N_-^2}{\Delta_b N_-^2} \equiv \frac{p^2}{p^2}\cdot \frac{ \Delta N_-^2}{ N_+}=\frac{\Delta S_-^2}{p S_+}.
    \ee
    In order to evaluate the average number of photons $p$ scattered by each atom, we compare fluorescence images to absorption images for increasingly larger atomic clouds~\cite{pigneurnon}. 
    From this comparison, we derived $p=29.4$ for the separation data and $p=20.7$ for the interference data, meaning each atom is generating, on average, clusters of around 20-30 photons when crossing the light-sheet. 
    We come now to the discussion on detection noise~\cite{bucker2013twin}.
    The final number of counts created by each photon hitting the camera is a random variable, whose statistics is governed by photonic shot noise.
    On top of the usual shot noise level, there is an additional noise generated by the amplification stage at the electron-multiplication register of the camera. To account for it, the variance due to shot noise gets doubled \cite{basden2003photon}: \(\Delta_{sn}S_-^2= 2S_+\).
    A second contribution comes from the background signal \(\hat{b}\) contained in a certain area of the camera chip, which is important when regions with low signal are considered. Since the background signal is indistinguishable from the actual signal coming from the atomic fluorescence, the same considerations made above apply and \(\Delta_{sn} b_-^2 = 2b_+\). 
    We define the total noise contribution to the variance $\Delta_n S_-^2 = \Delta_{sn} S_-^2+\Delta_{sn} b_-^2$ and modify the expression for the uncorrected number squeezing to take into account the total noise as
    \be
        \xi^2 = \frac{\Delta S_-^2 - \Delta_n S_-^2}{\Delta_b S_-^2}.
    \ee
    We can then define the minimum value of atom number squeezing $\xi^2_n$ between the momentum states detectable in our system as 
    \ba
        \xi^2_n = \frac{\Delta_n S_-^2}{\Delta_b S_-^2}=\frac{2S_+ + 2b_+}{pS_+} \simeq 2/p,\non \\~\text{for}~b_+ \ll S_+.
    \ea
    Typical values are $\xi^2_n \simeq 0.08$ (separation data) and $\xi^2_n \simeq0.2$ (interference data). The difference can be explained by the different signal-to-noise ratio $S_+/b_+$ for the two datasets.

\section{One-dimensional fit of the second-order correlation function}
    The one-dimensional fringe pattern of $g^{(2)}_{exp}(k^y_-,k^y_+)$ is fitted using the fit-function
    \begin{equation}\label{eq:fit_function}
        f(k_y)= \bigg[ d + C \cos(2\pi \frac{k_y-K}{e_{exp}}) \bigg]  \exp \bigg[\frac{-(k_y-K)^2 }{(c_{sigma}/e_{exp})^2}\bigg],
    \end{equation}
    where $K$ is the coordinate of the centre of the fringe pattern, $c_{sigma}$ is a dimensional parameter, $e_{exp}$ represents the diagonal fringe spacing, $C = 0.032 \pm 0.004$ is the contrast of the fringe pattern and $d$ an offset. 

\bibliographystyle{apsrev4-1}
\bibliography{SMb}